\documentclass[12pt,a4wide]{article}
\usepackage{amsmath,amssymb,amsfonts}

\newcommand{\eqa}{\begin{eqnarray}}
\newcommand{\ena}{\end{eqnarray}}
\newcommand{\topstar}[1]{\setlength{\unitlength}{1mm}
\begin{picture}(2,0)(-1,-1.4)
   \put(0,0){\makebox(0,0){$#1$}}
   \put(0,2.4){\makebox(0,0){\mbox{\tiny$\star$}}}
\end{picture}}
\begin{document}
\begin{center}
{\large {\bf Quasi-Metric Relativity}}
\end{center}
\begin{center}
Dag {\O}stvang \\
{\em Department of Physics, Norwegian University of Science and Technology
(NTNU) \\ N-7491 Trondheim, Norway}
\end{center}
\begin{abstract}
This is a survey of a new type of relativistic space-time framework;
the so-called quasi-metric framework. The basic geometric structure underlying
quasi-metric relativity is quasi-metric space-time; this is defined as a 
4-dimensional differentiable manifold ${\cal N}$ equipped with two 
one-parameter families ${\bf {\bar g}}_t$ and ${\bf g}_t$ of Lorentzian 
4-metrics parametrized by a global time function $t$. The ``dynamical'' metric 
family ${\bf {\bar g}}_t$ is found from field equations, whereas the 
``physical'' metric family ${\bf g}_t$ is used to propagate sources and to 
compare predictions to astronomical observations. A linear and symmetric affine 
connection ${\,}{\topstar {\nabla}}{\,}$ compatible with the family ${\bf g}_t$
is defined, giving rise to equations of motion.

Furthermore a quasi-metric theory of gravity, including field equations and
local conservation laws, is presented. Just as for General Relativity, the 
field equations accommodate two independent propagating dynamical degrees of 
freedom. On the other hand, the particular structure of quasi-metric geometry
allows only a partial coupling of space-time geometry to the active 
stress-energy tensor. Besides, the field equations are defined from 
projections of physical and geometrical tensors with respect to a 
``preferred'' foliation of quasi-metric space-time into spatial hypersurfaces. 
A number of non-standard features make these field equations unsuitable 
for a standard PPN-analysis, however. This implies that the experimental status
of the theory is not completely clear at this point in time. The theory seems 
to be consistent with a number of cosmological observations and it satisfies 
all the classical solar system tests, though. Moreover, in its non-metric 
sector the new theory has experimental support where General Relativity fails 
or is irrelevant.
\end{abstract}
\topmargin 0pt
\oddsidemargin 5mm
\section{Introduction} 
Interest in alternative classical theories of gravity has mainly focused
on the class of metric theories, defined by the postulates [1]
\begin{itemize}
{\item Space-time is equipped with a single Lorentzian metric field 
${\bf g}$,}
{\item The world lines of inertial test particles are geodesics of 
${\bf g}$,}
{\item In the local Lorentz frames, the non-gravitational physics is as 
in Special Relativity.}
\end{itemize}
One reason for the neglect of non-metric theories is probably the successes of
the leading metric theory, General Relativity (GR): constructing alternative 
theories not deviating too significantly in structure from GR seems compelling
if one is not prepared to risk immediate conflict with observations. 

But another reason is possibly the belief that theories which do not satisfy
the above postulates necessarily fail to satisfy the Einstein Equivalence
Principle (EEP) defined by the restrictions [1]
\begin{itemize}
{\item The trajectories of uncharged test particles do not depend on their
internal composition (this is the Weak Equivalence Principle),}
{\item The outcomes of local non-gravitational test experiments do not 
depend on the velocity of the apparatus (this is called local Lorentz 
invariance),}
{\item The outcomes of local non-gravitational test experiments do not 
depend on when or where they are performed (this is called local position
invariance).}
\end{itemize}
Since the empirical evidence supporting the EEP seems formidable [1],
constructing a theory violating it probably would be a waste of time. But is
it really true that theories not satisfying all the said postulates 
necessarily violate the EEP?

No, it is not. It can be shown that it is possible to construct a type of 
relativistic space-time framework not satisfying the first two postulates but 
where the EEP still holds [2]. This framework defines the geometrical 
basis for quasi-metric relativity (QMR), just as pseudo-Riemannian geometry 
defines the geometrical basis for metric relativity. 

A general physical motivation for introducing the quasi-metric framework is 
found directly in the particular global structure of quasi-metric space-time. 
That is, the geometric structure behind QMR is constructed to yield maximal 
predictive power with regard to the large-scale properties of space-time. The 
basic idea that makes this possible is that since the Universe is unique, so 
should the nature of its global evolution be. That is, there should be no 
reason to treat the Universe as a purely gravitational dynamical system and 
its global evolution should not depend on any particular choice of initial 
conditions. This means that the global evolution of the Universe should be 
explicitly included into the geometric structure of quasi-metric space-time as 
some sort of prior-geometric property. It is natural to call the global 
evolution of the Universe ``non-kinematical'' since by construction, this 
evolution is not part of space-time's causal structure and is unaffected by 
dynamics. In other words, in QMR the global evolution of the Universe is 
described as a non-kinematical cosmic expansion. One important consequence of 
this is that a global arrow of time exists as an intrinsic, geometrical 
property of quasi-metric space-time. The quasi-metric framework thus represents
an attractive solution of the problem of time-asymmetry. (See e.g. [3] and 
references therein for more on this problem.)

The quasi-metric framework and some of its predictions will be described in 
some detail in the following. Note that since this paper is intended to be a 
not too lengthy introduction to QMR, derivations of formulae are in general 
omitted. However, more detailed derivations can often be found in [2].
\section{The quasi-metric space-time framework} 
\subsection{Basic mathematical structure} 
As mentioned in the introduction, the basic premise of the quasi-metric 
framework is that the cosmic expansion should be described as a 
non-kinematical phenomenon. To fulfil this premise, it is necessary that the
canonical description of space-time is taken as fundamental. Furthermore, to 
ensure that the cosmic expansion really is global and directly integrated into
the geometric structure of quasi-metric space-time, it is necessary to 
introduce {\em the global time function} $t$ representing an extra, degenerate
time dimension (see below). This extra time dimension must be degenerate since
it is designed to describe the cosmic expansion as independent of space-time's
causal structure and taking the form of a non-kinematical global scale change 
between gravitational and non-gravitational systems. We will elaborate on this 
in the following where we define the quasi-metric framework precisely in terms 
of geometrical structures on a differentiable manifold.

Mathematically, the quasi-metric framework can be described by first 
considering a 5-dimensional product manifold ${\cal M}{\times}{\bf R}_1$, 
where ${\cal M}={\cal S}{\times}{\bf R}_2$ is a (globally hyperbolic)
Lorentzian space-time manifold, ${\bf R}_1$ and ${\bf R}_2$ are two copies of 
the real line and ${\cal S}$ is a compact Riemannian 3-dimensional manifold 
(without boundaries). It is natural to interpret $t$ as a 
coordinate on ${\bf R}_1$. Besides, the product topology of ${\cal M}$ implies
that once $t$ is given, there must exist a ``preferred'' ordinary time 
coordinate $x^0$ on ${\bf R}_2$ such that $x^0$ scales like $ct$. It is very 
convenient to choose a time coordinate $x^0$ which scales like $ct$ since
this means that $x^0$ is in some sense a mirror of $t$ and that they thus 
are ``equivalent'' global time coordinates but designed to parametrize
fundamentally different phenomena. A coordinate system with a global 
time coordinate of this type we call {\em a global time coordinate system} 
(GTCS). Hence, expressed in a GTCS ${\{}x^{\mu}{\}}$ (where ${\mu}$ can take 
any value $0-3$), $x^0$ is interpreted as a global time coordinate on 
${\bf R}_2$ and ${\{}x^j{\}}$ (where $j$ can take any value $1-3$) as spatial 
coordinates on ${\cal S}$. The class of GTCSs is a set of preferred coordinate
systems inasmuch as the equations of QMR take special forms in a GTCS. 

We now equip ${\cal M}{\times}{\bf R}_1$ with two degenerate 5-dimensional
metrics ${\bf {\bar g}}_t$ and ${\bf g}_t$. By definition, the ``dynamical'' 
metric ${\bf {\bar g}}_t$ represents a solution of field equations, and from 
${\bf {\bar g}}_t$ one can construct the ``physical'' metric ${\bf g}_t$ which
is used when comparing predictions to observations involving equations of 
motion. To reduce the 5-dimensional space-time ${\cal M}{\times}{\bf R}_1$ to 
a 4-dimensional space-time, we just slice the 4-dimensional sub-manifold 
${\cal N}$ determined by the equation $x^0=ct$ (using a GTCS) out of 
${\cal M}{\times}{\bf R}_1$. Moreover, in ${\cal N}$, ${\bf {\bar g}}_t$ and 
${\bf g}_t$ are interpreted as one-parameter metric families (this is merely a 
matter of semantics). Thus by construction, ${\cal N}$ is a 4-dimensional 
space-time manifold equipped with two one-parameter families of Lorentzian 
4-metrics parametrized by the global time function $t$. This is the general 
form of the quasi-metric space-time framework. We will call ${\cal N}$ 
{\em a quasi-metric space-time manifold.} And the reason why ${\cal N}$ is 
different from a Lorentzian space-time manifold is that the affine connection 
compatible with any metric family is non-metric. This means that, while it is 
always possible to equip ${\cal N}$ with the single metric obtained by 
inserting the explicit substitution $t=x^0/c$ into ${\bf g}_t$, this metric is 
useless for other purposes than taking scalar products. That is, since the 
affine structure on ${\cal N}$ is inherited from the affine structure on 
${\cal M}{\times}{\bf R}_1$, and since that affine structure is not compatible 
with any single metric on ${\cal N}$ (see below), one {\em must} separate 
between $ct$ and $x^0$ in ${\bf g}_t$.

From the definition of quasi-metric space-time, we see that it is constructed 
as consisting of two mutually orthogonal foliations: on the one hand 
space-time can be sliced up globally into a family of 3-dimensional space-like 
hypersurfaces (called the fundamental hypersurfaces (FHSs)) by the global 
time function $t$, on the other hand space-time can be foliated into a family 
of time-like curves everywhere orthogonal to the FHSs. These curves represent 
the world lines of a family of hypothetical observers called the fundamental 
observers (FOs), and the FHSs together with $t$ represent a preferred notion 
of space and time. That is, the equations of any theory of gravity based on 
quasi-metric geometry should depend on quantities obtained from this preferred
way of splitting up space-time into space and time. But notice that the extra
structure of quasi-metric space-time (as compared to Lorentzian space-time) 
has no effects on local non-gravitational test experiments.

Next we describe the affine structure on $({\cal N},{\bf g}_t)$. (Note that we 
introduce the coordinate notation $g_{(t){\mu}{\nu}}$ where the parenthesis is 
put in to emphasize that these are the components of a one-parameter family of 
metrics rather than those of a single metric.) Again we start with the 
corresponding structure on ${\cal M}{\times}{\bf R}_1$. To find that, we should 
think of the metric family ${\bf g}_t$ as one single degenerate metric on 
${\cal M}{\times}{\bf R}_1$, where the degeneracy manifests itself via the 
condition ${\bf g}_t({\frac{\partial}{{\partial}t}},{\cdot}){\equiv}0$. The 
natural way to proceed is to determine a torsion-free, metric-compatible 
5-dimensional ``degenerate'' connection 
${\stackrel{{\hbox{\tiny$\star$}}}{\nabla}}$ on ${\cal M}{\times}{\bf R}_1$ 
from the metric-preserving condition
\eqa
{\frac{\partial}{{\partial}t}}{\bf g}_t({\bf y}_t,
{\bf z}_t)={\bf g}_t({\stackrel{{\hbox{\tiny$\star$}}}{\nabla}}_{{\!}{\!}
{\frac{\partial}{{\partial}t}}}{\bf y}_t,{\bf z}_t)+{\bf g}_t({\bf y}_t,
{\stackrel{{\hbox{\tiny$\star$}}}{\nabla}}_{{\!}{\!}
{\frac{\partial}{{\partial}t}}}{\bf z}_t),
\ena
involving arbitrary families of vector fields ${\bf y}_t$ and ${\bf z}_t$ in 
${\cal M}$. It turns out that it is possible to find a unique candidate 
connection satisfying equation (1) in general and differing from the usual 
Levi-Civita connection only via connection coefficients containing $t$.
This candidate connection is determined from the in general non-zero connection
coefficients ${\topstar {\Gamma}}^{{\,}{\alpha}}_{{\mu}t}$ which must be equal to 
${\frac{1}{2}}g_{(t)}^{{\alpha}{\sigma}}
{\frac{\partial}{{\partial}t}}g_{(t){\sigma}{\mu}}$ (we use Einstein's
summation convention throughout), since other connection coefficients 
containing $t$ must vanish identically.

But the above-mentioned candidate degenerate connection has one undesirable
property, namely that it does not in general ensure that the unit normal 
vector field family ${\bf n}_t$ of the FHSs (with the property 
${\bf g}_t({\bf n}_t,{\bf n}_t)=-1$) is parallel-transported along 
${\frac{\partial}{{\partial}t}}$. It would be natural to require this 
property, i.e., ${\stackrel{{\hbox{\tiny$\star$}}}{\nabla}}$ should guarantee 
that
\eqa
{\stackrel{{\hbox{\tiny$\star$}}}{\nabla}}_{{\!}{\!}
{\frac{\partial}{{\partial}t}}}{\bf n}_t=0,
\ena 
since if equation (2) does not hold the resulting equations of motion will 
not be identical to the geodesic equation derived from
${\stackrel{{\hbox{\tiny$\star$}}}{\nabla}}$. However, we may try to construct
a unique degenerate connection which satisfies equation (2) and is identical to
the above-mentioned candidate connection for those particular cases when 
the candidate connection satisfies equation (2). This is possible since the 
dependence of ${\bf g}_t$ on $t$ cannot be arbitrary. That is, the (global) 
explicit dependence of ${\bf g}_t$ and ${\bf {\bar g}}_t$ on $t$ is the same
and it can be inferred independently. Moreover, it takes a particular form 
(see equation (13) below), making it possible to construct a unique degenerate 
connection which satisfies both equations (1) and (2) (given the particular 
dependence of ${\bf g}_t$ on $t$). This unique connection is determined from 
the form the connection coefficients take in a GTCS, and it involves the 
components $h_{(t)ij}$ of the family ${\bf h}_t$ of spatial metrics intrinsic 
to the FHSs. Said (nonzero) connection coefficients read
\eqa 
{\topstar {\Gamma}}^{{\,}i}_{jt}={\frac{1}{2}}h_{(t)}^{is}
{\frac{\partial}{{\partial}t}}h_{(t)sj}, {\qquad}
{\topstar {\Gamma}}^{{\,}i}_{tj}{\,}{\equiv}
{\topstar {\Gamma}}^{{\,}i}_{jt}, {\qquad}
{\topstar {\Gamma}}^{{\,}{\alpha}}_{{\nu}{\mu}}{\equiv}
{\Gamma}^{\alpha}_{(t){\nu}{\mu}},
\ena
where ${\Gamma}^{\alpha}_{(t){\nu}{\mu}}$ are the connection coefficients of 
the family ${\nabla}_t$ of Levi-Civita connections defined from the 
collection of single Lorentzian metrics on ${\cal M}$. The restriction of 
${\stackrel{{\hbox{\tiny$\star$}}}{\nabla}}$ to ${\cal N}$ is trivial since it
does not involve any projections. (That is, to apply 
${\stackrel{{\hbox{\tiny$\star$}}}{\nabla}}$ in $\cal N$ one just applies it 
in the sub-manifold $x^0=ct$ in a GTCS.) Observe that other degenerate 
connection coefficients than those given in equation (3) vanish identically. 
This implies that the gradient of the global time function is covariantly 
constant, i.e., ${\stackrel{{\hbox{\tiny$\star$}}}{\nabla}}_{{\!}{\!}
{{\frac{\partial}{{\partial}t}}}}dt=
{\stackrel{{\hbox{\tiny$\star$}}}{\nabla}}_{{\!}{\!}
{{\frac{\partial}{{\partial}x^{\mu}}}}}dt=0$.

It is in general possible to write ${\bf g}_t$ as a sum of two terms
\eqa
{\bf g}_t=-{\bf g}_t({\bf n}_t,{\cdot}){\otimes}
{\bf g}_t({\bf n}_t,{\cdot})+{\bf h}_t.
\ena
Then equations (1) and (2) imply that
\eqa
{\stackrel{{\hbox{\tiny$\star$}}}{\nabla}}_{{\!}{\!}
{\frac{\partial}{{\partial}t}}}{\bf g}_t=0, \qquad
{\stackrel{{\hbox{\tiny$\star$}}}{\nabla}}_{{\!}{\!}
{\frac{\partial}{{\partial}t}}}{\bf h}_t=0,
\ena
thus the degenerate connection is compatible with (the non-degenerate part of)
the metric family ${\bf g}_t$ as asserted. 
\subsection{General equations of motion}
Now we want to use the above defined affine structure on ${\cal N}$ to find 
equations of motion for test particles in $({\cal N},{\bf g}_t)$. Let 
${\lambda}$ be an affine parameter along the world line in ${\cal N}$ of an 
arbitrary test particle. (In addition to the affine parameter $\lambda$, $t$ 
is also a (non-affine) parameter along any non-space-like curve in $\cal N$.) 
Using an arbitrary coordinate system (not necessarily a GTCS) we may define 
coordinate vector fields ${\frac{\partial}{{\partial}x^{\alpha}}}$; then
${\frac{dt}{d{\lambda}}}{\frac{\partial}{{\partial}t}}+
{\frac{dx^{\alpha}}{d{\lambda}}}{\frac{\partial}{{\partial}x^{\alpha}}}$ is 
the coordinate representation of the tangent vector field 
${\frac{\partial}{{\partial}{\lambda}}}$ along the curve. We then define the
degenerate covariant derivative along the curve as
\eqa
{\stackrel{{\hbox{\tiny$\star$}}}{\nabla}}_{{\!}{\!}
{\frac{\partial}{{\partial}{\lambda}}}}{\equiv}
{\frac{dt}{d{\lambda}}}{\stackrel{{\hbox{\tiny$\star$}}}{\nabla}}_{{\!}{\!}
{\frac{\partial}{{\partial}t}}}+{\frac{dx^{\alpha}}{d{\lambda}}}
{\stackrel{{\hbox{\tiny$\star$}}}{\nabla}}_{{\!}{\!}
{\frac{\partial}{{\partial}x^{\alpha}}}}.
\ena
A particularly important family of vector fields is the 4-velocity tangent 
vector field family ${\bf u}_t$ along a curve. By definition we have
\eqa
{\bf u}_t{\equiv}u^{\alpha}_{(t)}{\frac{\partial}{{\partial}x^{\alpha}}}
{\equiv}
{\frac{dx^{\alpha}}{d{\tau}_t}}{\frac{\partial}{{\partial}x^{\alpha}}},
\ena
where ${\tau}_t$ is the proper time as measured along the curve.

The equations of motion are found by calculating the covariant derivative of 
4-velocity tangent vectors along themselves using the connection in 
$({\cal N},{\bf g}_t)$. According to the above, this is equivalent to 
calculating ${\stackrel{{\hbox{\tiny$\star$}}}{\nabla}}{\bf u}_t$ along 
${\frac{\partial}{{\partial}{\tau}_t}}$. Using the coordinate 
representation of ${\frac{\partial}{{\partial}{\tau}_t}}$ we may thus 
define the vector field ${\stackrel{\star}{\bf a}}$ by
\eqa
{\stackrel{\star}{\bf a}}{\equiv}{\stackrel{{\hbox{\tiny$\star$}}}{\nabla}}
_{{\!}{\!}{\frac{\partial}{{\partial}{\tau}_t}}}{\bf u}_t={\Big (}
{\frac{dt}{d{\tau}_t}}{\stackrel{\star}{\nabla}}_{{\!}{\!}
{\frac{\partial}{{\partial}t}}}+{\frac{dx^{\alpha}}{d{\tau}_t}}
{\stackrel{{\hbox{\tiny$\star$}}}{\nabla}}_{{\!}{\!}
{\frac{\partial}{{\partial}x^{\alpha}}}}{\Big )}{\bf u}_t
{\equiv}{\frac{dt}{d{\tau}_t}}{\stackrel{{\hbox{\tiny$\star$}}}{\nabla}}
_{{\!}{\!}{\frac{\partial}{{\partial}t}}}{\bf u}_t + {\bf a}_t.
\ena
We call this vector field the ``degenerate" 4-acceleration. We need to have 
an independent expression for the degenerate acceleration field 
${\stackrel{\star}{{\bf a}}}$. This can be found by calculating the extra 
term ${\frac{dt}{d{\tau}_t}}{\stackrel{{\hbox{\tiny$\star$}}}{\nabla}}
_{{\!}{\!}{\frac{\partial}{{\partial}t}}}{\bf u}_t$ at the right hand side of 
equation (8). To do that, it is convenient to introduce the 3-velocity 
${\bf w}_t$ of an arbitrary test particle with respect to the FOs. That is, one 
may split up the tangent 4-velocity into parts respectively normal and 
tangential to the FHSs, i.e.
\eqa
{\bf u}_t={\stackrel{{\hbox{\tiny$\star$}}}{\gamma}}
(c{\bf n}_t+{\bf w}_t), \qquad
{\stackrel{{\hbox{\tiny$\star$}}}{\gamma}}
{\equiv}(1-{\frac{w^2}{c^2}})^{-1/2}={\frac{d{\tau}_{\cal F}}{d{\tau}_t}},
\ena
where $w^2$ is the square of ${\bf w}_t$ and $d{\tau}_{\cal F}{\equiv}Ndt$ is 
the proper time interval of the local FO. Here $N$ is the lapse function field
(as expressed in a GTCS) of the FOs. Note that ${\bf w}_t$ is an object 
intrinsic to the FHSs since ${\bf g}_t({\bf n}_t,{\bf w}_t){\equiv}0$. 
Moreover, from equation (5) we have that
${\stackrel{{\hbox{\tiny$\star$}}}{\nabla}}
_{{\!}{\!}{\frac{\partial}{{\partial}t}}}{\bf g}_t({\bf w}_t,{\bf w}_t)=0$, so $w^2$ is 
independent of $t$. This result, in combination with the
connection coefficients given in equation (3), yields that
$h_{(t)ij}{\stackrel{{\hbox{\tiny$\star$}}}{\nabla}}
_{{\!}{\!}{\frac{\partial}{{\partial}t}}}w_{(t)}^i=0$ and thus 
${\stackrel{{\hbox{\tiny$\star$}}}{\nabla}}
_{{\!}{\!}{\frac{\partial}{{\partial}t}}}{\bf w}_t=0$. This means, using equations (2) 
and (9), that we in fact have ${\stackrel{{\hbox{\tiny$\star$}}}{\nabla}}
_{{\!}{\!}{\frac{\partial}{{\partial}t}}}{\bf u}_t=0$ and thus
${\stackrel{\star}{\bf a}}$=${\bf a}_t$ from equation (8). The coordinate 
expression for ${\stackrel{\star}{\bf a}}$ then yields equations of motion, 
namely
\eqa
{\frac{d^2x^{\alpha}}{d{\lambda}^2}}+{\Big(}
{\topstar{\Gamma}}^{{\,}{\alpha}}_{t{\sigma}}{\frac{dt}{d{\lambda}}}+
{\topstar{\Gamma}}^{{\,}{\alpha}}_{{\beta}{\sigma}}
{\frac{dx^{\beta}}{d{\lambda}}}{\Big)}{\frac{dx^{\sigma}}{d{\lambda}}}=
{\Big(}{\frac{d{\tau}_t}{d{\lambda}}}{\Big)}^2a_{(t)}^{\alpha}.
\ena
Equation (10) is the geodesic equation obtained from 
${\stackrel{{\hbox{\tiny$\star$}}}{\nabla}}$ and this implies that inertial 
test particles follow geodesics of 
${\stackrel{{\hbox{\tiny$\star$}}}{\nabla}}$. Note that,
while the form of equation (10) is valid in general coordinates, the form of 
${\topstar {\Gamma}}^{{\,}{\alpha}}_{{\mu}t}$ given in equation (3) is not.

From the general global dependence of ${\bf {\bar g}}_t$ and ${\bf g}_t$ on
$t$ given in equation (13) below we can find the coordinate expressions in a 
GTCS of ${\bf n}_t$ and ${\bf w}_t$. These are given by [2]
\eqa
n^0_{(t)}=N^{-1}, \qquad n^j_{(t)}=-{\frac{t_0}{t}}{\frac{N^j_{(t)}}{N}}, 
\qquad w^0_{(t)}=0, \qquad w^j_{(t)}={\frac{dx^j}{d{\tau}_{\cal F}}}+
{\frac{t_0}{t}}{\frac{N^j_{(t)}}{N}}c,
\ena
where the ${\frac{t_0}{t}}N^j_{(t)}$ are the components in a GTCS 
of the shift vector field family entering the line element family 
${ds}_t^2$ representing $({\cal N},{\bf g}_t)$. (Here, $t_0$ is just some 
arbitrary epoch setting the scale of the spatial coordinates.) Now we firstly
notice that $N$ does not depend explicitly on $t$. Secondly, we notice that for
equation (2) to hold, we must have that
\eqa
{\frac{\partial}{{\partial}t}}{\Big (}{\frac{t^2_0}{t^2}}N^i_{(t)}
N^j_{(t)}h_{(t)ij}{\Big )}=0,
\ena
determining the dependence of $N^j_{(t)}$ on $t$. Thirdly, we notice that the 
proper time interval $d{\tau}_{\cal F}{\equiv}Ndt$ may in principle be 
integrated along the world line of any FO given the implicit dependence 
$N(x^{\mu}(t))$ in $({\cal N},{\bf g}_t)$. This means that there is a direct 
relationship between $t$ and the proper time elapsed for any FO. So, since $N$ 
is non-negative by definition, $t$ must be increasing in the forward direction 
of proper time for any FO.

To get the correspondence with metric gravity, we formally set 
${\frac{t_0}{t}}=1$ and then take the limit $t{\rightarrow}{\infty}$ in 
equations (3), (10) and (11). The equations of motion (10) then reduce to the 
usual geodesic equation in metric gravity. This limit represents the so-called 
{\em metric approximation} where the metric family ${\bf g}_t$ does not 
depend on $t$. That is, in the metric approximation, ${\bf g}_t$ can be 
identified with one single Lorentzian metric ${\bf g}$. Note that in QMR,
metric approximations are meaningful for isolated systems only. This is why
correspondences between QMR and  metric gravity can be found for, e.g., the 
solar system but not for cosmology (see section 4).

However, except for in the metric approximation, ${\bf g}_t$ should not be 
identified with any single Lorentzian metric and equation (10) does not reduce 
to the usual geodesic equation in metric gravity due to terms explicitly 
depending on $t$. That is, in QMR inertial test particles do not move as if 
they were following geodesics of any single space-time metric. Also note that 
the equations of motion (10) do not violate local Lorentz invariance. To see 
this, observe that the connection coefficients may be made to vanish in any 
local inertial frame so that equation (10) takes its special relativistic form.
\section{Quasi-metric gravity}
\subsection{Basic principles}
At this point two questions naturally arise, namely 
\begin{itemize}
{\item What is the role of $t$ in the metric families ${\bf g}_t$ and 
${\bf {\bar g}}_t$? and}
{\item Why should it be preferable to describe space-time 
by a metric family rather than by a single Lorentzian metric field?}
\end{itemize}
The answer to the first question should be clear from the discussion in the
previous sections. That is, by definition the main role of $t$ in the metric 
families is to describe global scale changes of the FHSs as measured by the 
FOs. This means that $t$ should enter each metric family explicitly as a 
spatial scale factor $R(t)$. To avoid introducing any extra arbitrary scale
or parameter we just define $R(t)=ct$. (Further justification of this choice
of scale factor will be given later in this section.) Moreover, the FHSs are by 
definition compact to ensure the uniqueness of a global time coordinate. That 
is, by requiring the FHSs to be compact, we ensure that $t$ splits quasi-metric 
space-time into a unique set of FHSs. Besides, since there is no reason to 
introduce any nontrivial spatial topology, the global basic geometry of the 
FHSs (neglecting the effects of gravity) should be that of the 3-sphere 
${\bf S}^3$. But any restriction on the global geometry of the FHSs implies 
the existence of prior 3-geometry. This prior 3-geometry should not restrict 
the general form of the metric family explicitly though; rather it should
be represented by specific terms in the field equations (see section 3.3).
 
We now set up the general form of the metric family 
$({\cal N},{\bf {\bar g}}_t)$, represented by the family of line elements 
${\overline {ds}}_t^2$, where the global dependence on $t$ is included 
explicitly via the scale factor mentioned above. That is, expressed in a 
suitable GTCS, the most general form allowed for the family ${\bf {\bar g}}_t$ 
may be represented by the family of line elements (we use the metric signature 
$(-+++)$ and Einstein's summation convention throughout)
\eqa
{\overline {ds}}_t^2={\bar N}_t^2{\Big \{ }
[{\bar N}_{(t)}^k{\bar N}_{(t)}^s{\tilde h}_{(t)ks}-1](dx^0)^2+
2{\frac{t}{t_0}}{\bar N}_{(t)}^k{\tilde h}_{(t)ks}dx^sdx^0+
{\frac{t^2}{t_0^2}}{\tilde h}_{(t)ks}dx^kdx^s{\Big \} }.
\ena
Here, $t_0$ is some arbitrary reference epoch (usually chosen to be the present
epoch) setting the scale of the spatial coordinates, ${\bar N}_t$ is the family
of lapse functions (of the FOs) and ${\frac{t_0}{t}}{\bar N}^k_{(t)}$ are the 
components of the shift vector family (of the FOs) in 
$({\cal N},{\bf {\bar g}}_t)$. Also, ${\bar h}_{(t)ks}{\equiv}{\frac{t^2}{t_0^2}}
{\bar N}_t^2{\tilde h}_{(t)ks}$ are the components of the spatial metric family 
${\bf {\bar h}}_t$ intrinsic to the FHSs. We notice that ${\bar N}_t^2$ enters 
equation (13) as a conformal factor, but that this does not imply any 
restrictions on the general form of ${\bf {\bar g}}_t$. The only point of 
introducing the conformal factor is that the total scale factor of the FHSs may
now be defined formally as ${\bar F}_t{\equiv}{\bar N}_tct$. We also notice 
that ${\bar N}_t$ may depend on $t$ and that the form (13) of 
${\bf {\bar g}}_t$ is preserved only under coordinate transformations between 
GTCSs. Furthermore, we notice that the most general allowed metric 
approximation of ${\bf {\bar g}}_t$ is the single metric ${\bf {\bar g}}$ 
obtained from equation (13) by setting ${\frac{t}{t_0}}=1$ and eliminating all 
remaining $t$-dependence by making the substitution $t=x^0/c$. Finally, as we 
shall see later, we notice that the field equations will impose restrictions on
the family of space line elements 
$d{\tilde{\sigma}}_t^2{\equiv}{\tilde h}_{(t)ks}dx^kdx^s$ (representing the 
space metric family ${\bf {\tilde h}}_t$ conformal to ${\bf {\bar h}}_t$) 
entering equation (13). This means that the metric family (13) will contain 
prior 3-geometry, so that the form of ${\bf {\bar g}}_t$ will be somewhat less 
general than might be expected from equation (13).

As mentioned earlier, to get the correct affine structure on $({\cal N},
{\bf g}_t)$ one {\em must} separate between $ct$ and $x^0$ in ${\bf g}_t$. 
Similarly, to get the correct affine structure on $({\cal N},
{\bf {\bar g}}_t)$, one must separate between $ct$ and $x^0$ in equation (13).
But the possibility that ${\bar N}_t$ depends explicitly on $t$ means that the
affine structure on $({\cal N},{\bf {\bar g}}_t)$ will differ slightly from 
that on $({\cal N},{\bf g}_t)$. That is, since counterparts to equations (2) 
and (5) must be valid in $({\cal N},{\bf {\bar g}}_t)$ as well, i.e.,
\eqa
{\stackrel{{\hbox{\tiny$\star$}}}{\bf {\bar {\nabla}}}}_{{\!}{\!}
{\frac{\partial}{{\partial}t}}}{\bf {\bar g}}_t=0, \qquad
{\stackrel{{\hbox{\tiny$\star$}}}{\bf {\bar {\nabla}}}}_{{\!}{\!}
{\frac{\partial}{{\partial}t}}}{\bf {\bar n}}_t=0, \qquad
{\stackrel{{\hbox{\tiny$\star$}}}{\bf {\bar {\nabla}}}}_{{\!}{\!}
{\frac{\partial}{{\partial}t}}}{\bf {\bar h}}_t=0,
\ena
the potential $t$-dependence of ${\bar N}_t$ implies that the degenerate 
connection coefficients in $({\cal N},{\bf {\bar g}}_t)$ will not take a form 
exactly like that shown in equation (3). Rather, the in general non-vanishing 
connection coefficients in $({\cal N},{\bf {\bar g}}_t)$ are given in a GTCS 
by (a comma denotes taking a partial derivative)
\eqa
{\topstar{\bar {\Gamma}}}^{{\,}0}_{t0}={\frac{{\bar N}_t,_t}{{\bar N}_t}}, 
\quad {\topstar{\bar {\Gamma}}}_{tj}^{{\,}i}=
{\Big (}{\frac{1}{t}}+{\frac{{\bar N}_{t,t}}{{\bar N}_t}}{\Big )}{\delta}^i_j
+{\frac{1}{2}}{\tilde h}_{(t)}^{is}{\tilde h}_{(t)sj,t}, \quad
{\topstar{\bar {\Gamma}}}^{{\,}{\alpha}}_{t{\mu}}{\equiv}
{\topstar{\bar {\Gamma}}}^{{\,}{\alpha}}_{{\mu}t}, \qquad
{\topstar{\bar {\Gamma}}}^{{\,}{\alpha}}_{{\nu}{\mu}}{\equiv}
{\bar {\Gamma}}^{\alpha}_{(t){\nu}{\mu}}.
\ena
Note that for equation (14) to hold, we must have (this is a counterpart 
to equation (12))
\eqa
{\frac{\partial}{{\partial}t}}{\Big (}{\bar N}^i_{(t)}
{\bar N}^j_{(t)}{\tilde h}_{(t)ij}{\Big )}=0.
\ena
Next, we want to describe the evolution of the total spatial scale factor 
${\bar F}_t$ of the FHSs in the hypersurface-orthogonal 
direction. By definition we have (the symbol `${\bar {\perp}}$' denotes a 
scalar product with $-{\bf {\bar n}}_t$)
\eqa
{\bar F}_t^{-1}{\topstar{\cal L}}_{{\bf {\bar n}}_t}{\bar F}_t{\equiv}
{\bar F}_t^{-1}{\Big (}(c{\bar N}_t)^{-1}{\bar F}_t,_t+
{\cal L}_{{\bf {\bar n}}_t}{\bar F}_t{\Big )}=
(c{\bar N}_t)^{-1}{\Big [}{\frac{1}{t}}+{\frac{{\bar N}_t,_t}{{\bar N}_t}}
{\Big ]}-{\frac{{\bar N}_t,_{\bar {\perp}}}{{\bar N}_t}}{\equiv}
c^{-2}{\bar x}_t+c^{-1}{\bar H}_t,
\ena
where ${\cal L}_{{\bf {\bar n}}_t}$ denotes a projected Lie derivative (operating 
on objects intrinsic to the FHSs only) in the direction normal to the FHSs, 
treating $t$ as a constant where it occurs explicitly. In equation (17), 
$c^{-2}{\bar x}_t$ represents the {\em kinematical contribution} to the 
evolution of ${\bar F}_t$ and $c^{-1}{\bar H}_t$ represents the so-called 
{\em non-kinematical contribution} defined by
\eqa
{\bar H}_t{\equiv}{\frac{1}{{\bar N}_tt}}+{\bar y}_t, \qquad
{\bar y}_t{\equiv}c^{-1}{\sqrt{{\bar a}_{{\cal F}k}{\bar a}_{\cal F}^k}},
\qquad c^{-2}{\bar a}_{{\cal F}j}{\equiv}{\frac{{\bar N}_t,_j}{{\bar N}_t}},
\ena
where ${\bf {\bar a}}_{\cal F}$ is the 4-acceleration field of the FOs in the 
family ${\bf {\bar g}}_t$. We see that the non-kinematical evolution (NKE) of 
the spatial scale factor takes the form of an ``expansion'' since ${\bar H}_t$
can never take negative values. Furthermore, we observe that ${\bar H}_t$ does 
not vanish even if the kinematical evolution of ${\bar F}_t$ does and 
${\bar N}_t$ is a constant in $({\cal N},{\bf {\bar g}}_t)$. For this 
particular case (see equation (57) below) we have the relationship 
${\bar H}_t={\frac{1}{{\bar N}_tt}}={\sqrt {\frac{{\bar P}_t}{6}}}c$, where 
${\bar P}_t$ is the Ricci scalar curvature family intrinsic to the FHSs. This 
means that in quasi-metric relativity, a global increase in scale of the FHSs
is linked to their global curvature. Moreover, this global increase of scale 
has nothing to do with the kinematical structure described by any single 
Lorentzian metric field. It follows that in QMR, the Hubble law is not 
interpreted as a kinematical law, rather the Hubble law is interpreted as 
evidence for global spatial curvature. This reinterpretation of the Hubble law
also justifies the choice of a scale factor ${\bar F}_t{\propto}t$ since no 
other choice fulfils the above relationship with ${\bar H}_t$ playing the 
role of a ``Hubble parameter'' for the special case when ${\bar N}_t$ is a 
constant. In particular, note that it is not possible to construct similar 
models where the global NKE takes the form of a ``contraction'' without 
introducing some extra arbitrary scale.

It follows from the above discussion that quasi-metric space-time is 
manifestly time-asymmetric by construction, irrespective of the fact that 
dynamical laws are time-reversal invariant. That is, quasi-metric space-time 
is time-asymmetric regardless of whether solutions of dynamical equations are 
time-symmetric or not. For example, one may find time-symmetric (e.g., static) 
solutions for ${\bar N}_t$ in equation (13). But the scale factor is never 
time-symmetric, as can be seen from equation (17). This illustrates that the 
global time-asymmetry of quasi-metric space-time is due to the cosmological 
arrow of time represented by the global cosmic expansion. Moreover, the 
existence of a global arrow of time means that quasi-metric space-time has
a simple causal structure.

The answer to the second question we posed above is now clear. In order to 
construct a new theory with considerably more predictive power than metric 
theory in cosmology, it would seem necessary to describe the cosmic expansion 
as non-kinematical, i.e., as some sort of prior-geometric property of 
space-time itself. In this way an enormous multitude of possibilities 
regarding cosmic genesis and evolution will be eliminated just by
postulating that space-time is quasi-metric. This is why quasi-metric 
space-time should be preferred over Lorentzian space-time as a matter of
principle as long as this position is not in conflict with observations.
\subsection{Units and measurement}
The fact that QMR describes global scale changes of the FHSs as 
non-kinematical suggests the existence of two fundamentally different scales 
in the Universe, one gravitational and one atomic. This means that we have to 
specify which kind of units we are supposed to use in equation (13). In metric
theory it does not matter which kind of units one uses, but in quasi-metric 
theory this is not so obvious. That is, is equation (13) equally valid in 
units operationally defined from systems where gravitational interactions are 
dominant, as in operationally defined atomic units based on systems where 
gravitational interactions are insignificant? It turns out that the answer to 
this question is negative.

The units implicitly assumed when writing down line elements of the type (13)
should be ``atomic'' units; i.e., units operationally defined by using atomic 
clocks and rods only. This means that we may interpret the variation in 
space-time of the spatial scale factor ${\bar F}_t$ as a consequence of the
fact that we use atomic units to measure gravitational scales. Equivalently,
we may interpret the variation of ${\bar F}_t$ to mean that by definition,
operationally defined atomic units are considered formally variable throughout
space-time. (This interpretation is possible since any non-local 
inter-comparison of operationally defined units is purely a matter of 
definition.) The formal variation of atomic units in space-time means that 
gravitational quantities get an extra formal variation when measured in atomic
units (and {\em vice versa}). (This will show up explicitly, e.g., in 
differential laws such as local conservation laws.) We now postulate that 
atomic length units vary in space-time just as the inverse of the spatial scale
factor ${\bar F}_t$ since this implies that the scale of the FHSs does not vary 
measured in gravitational units. That is, any gravitational quantity gets a 
formal variability as some power of ${\bar F}_t$ when measured in atomic units.
By definition $c$ and Planck's constant ${\hbar}$ are not formally variable 
(this yields no physical restrictions since, in the absence of other 
dimensionful quantities, $c$ and ${\hbar}$ cannot be combined to 
get a dimensionless number). This means that the formal variation of atomic 
length and atomic time units are identical and inverse to that of atomic energy
(or mass) units (charge units have no formal variation). One particular 
consequence of this formal variability is that the ``bare'' gravitational 
coupling parameter $G_t^{\rm B}$ must be variable when measured in atomic units. 

By dimensional analysis, it is found that $G_t^{\rm B}$ varies like
coordinate length squared measured in atomic units (i.e., as ${\bar F}_t^2$). 
Besides, $G_t^{\rm B}$ couples to charge squared, or more generally to the 
electromagnetic stress-energy tensor [4]. On the other hand, for material 
sources, masses formally vary as ${\bar F}_t^{-1}$ but this is not measurable in 
non-gravitational experiments. This means that the ``screened'' gravitational 
coupling parameter $G_t^{\rm S}$ measured for material sources effectively 
varies as ${\bar F}_t$. Since $G_t^{\rm B}$ and $G_t^{\rm S}$ usually occur in 
combination with charge or mass, it is convenient to define $G_t^{\rm B}$ and 
$G_t^{\rm S}$ to take the constant values $G^{\rm B}$ and $G^{\rm S}$, 
respectively, measured in (hypothetical) local gravitational experiments in an 
empty universe at epoch $t_0$. But if one does this, one must separate between 
{\em active mass} $m_t$ measured dynamically as a source of gravity and 
{\em passive mass m} (i.e., passive gravitational mass or inertial mass). 
(Similarly one must separate between {\em active charge} and {\em passive 
charge} [4].) That is, we include the formal variation of $G_t^{\rm B}$ and 
$G_t^{\rm S}$ into $m_t$. This means that the formal variation of active mass 
goes as ${\bar F}_t$ for any material particle, whereas for the electromagnetic 
field, the formal variation of active mass (or energy) goes as ${\bar F}_t^2$.
That is, active masses vary locally as 
\eqa
m_t=
\left\{
\begin{array}{ll}
{\frac{{\bar N}_tt}{{\bar N}_0t_0}}m_0, 
& \qquad $$\hbox{for material particles,}$$ \\[1.5ex]
{\frac{{\bar N}_t^2t^2}{{\bar N}^2_0t_0^2}}m_0,, 
& \qquad $$\hbox{for the electromagnetic field,}$$
\end{array}
\right.
\ena
where ${\bar N}_0$ and $m_0$ denote values at some arbitrary reference event. 
(Formal variations of other gravitational quantities may be found similarly.)
By convention we choose ${\bar N}_0=1$; this means that the (hypothetical) 
reference situation is an empty Universe at epoch $t_0$ (see section 4).
A consequence of equation (19) is that local gravitational experiments designed
to measure gravitational coupling parameters should depend on source 
composition, so that it will be necessary to distinguish between $G^{\rm B}$ 
and $G^{\rm S}$.

However, the electromagnetic field also in principle experiences a cosmic
``redshift''-effect (not noticeable locally) and also gravitational spectral 
shifts [4]; these spectral shifts yield an extra factor ${\bar F}_t^{-1}$. So, 
if an extended gravitational source is expanding (as for, e.g., a 
gravitationally bound source made of ideal gas [5]), one must take into account
both the formal variation of active mass and the cosmic ``redshift'' of the 
electromagnetic field. For this case, the variation due to both said effects
of the total active stress-energy tensor ${\bf T}_t$ (considered as a source of
gravitation) goes as ${\bar F}_t^{-2}$ if one can neglect net energy transfer 
between the electromagnetic field and material particles. (But in general there
will be thermal contact between photons and material particles, so that the 
cosmic expansion cools the photons again cooling the material particles via 
thermal contact. This means that the active electromagnetic stress-energy 
tensor ${\bf T}^{\rm (EM)}_t$  will decrease a little more slowly than 
${\frac{t^2_0}{t^2}}$ and that the active material particle stress-energy 
tensor ${\bf T}^{\rm (MA)}_t$ will decrease a little faster than 
${\frac{t^2_0}{t^2}}$.) On the other hand, if the gravitational source does not
expand (as for, e.g., a solid hollow cavity filled with ideal gas), the fact 
that material sources and the electromagnetic field have different formal 
variations of active mass-energy means that ${\bf T}^{\rm (EM)}_t$ and
${\bf T}^{\rm (MA)}_t$ will have very different secular evolutions.

Note that the necessity to separate between gravitational and atomic scales 
represents a violation of the Strong Equivalence Principle (SEP). Also observe 
that, since ${\bar F}_t$ is a constant in the Newtonian limit, Newtonian theory
with a formally variable $G_t^{\rm S}$ (or $G_t^{\rm B}$) will be inconsistent 
with the Newtonian limit of QMR.
\subsection{Field equations}
Now we must find field equations which represent couplings between 
space-time geometry and the active stress-energy tensors ${\bf T}^{\rm (EM)}_t$ 
and ${\bf T}^{\rm (MA)}_t$ for the electromagnetic field and material sources, 
with the two different coupling constants $G^{\rm B}$ and $G^{\rm S}$, 
respectively. It would seem natural to look for field equations with structure
somewhat similar to that of the Einstein field equations. However, apart from
the non-universal gravitational coupling, there should also be other 
significant differences. 

First, QMR describes the time evolution of one distinguished foliation of 
space-time into spatial hypersurfaces, i.e., the FHSs. The field equations 
must be tailored to this particular foliation and not to alternative 
foliations. In fact, the distinguished foliation should be determined as part 
of any solution of the field equations. But this means that the field equations
cannot involve ``genuine'' space-time tensors only; also some geometric tensor 
stemming from the distinguished foliation must be introduced. Second, in 
contrast to GR, in quasi-metric gravity we have so-called ``prior'' geometry, 
i.e., non-dynamical aspects of the space-time geometry which are not influenced
by matter sources. This is so since we have restricted the metric family 
${\bf {\bar g}}_t$ by requiring that the ``basic'' geometry of the FHSs should 
be that of the 3-sphere. This prior geometry should be represented by specific 
terms in the field equations. 

Fortunately, it is possible to avoid the above concerns for a subset of the 
field equations, yielding a metric correspondence with GR (and Newtonian 
theory), in a natural, geometrical way. That is, we postulate one field 
equation valid for projections with respect to the FHSs, namely (using a GTCS)
\eqa
2{\bar R}_{(t){\bar {\perp}}{\bar {\perp}}}=
2(c^{-2}{\bar a}_{{\cal F}{\mid}s}^s+
c^{-4}{\bar a}_{{\cal F}s}{\bar a}_{\cal F}^s-{\bar K}_{(t)ks}
{\bar K}_{(t)}^{ks}+{\cal L}_{{\bf {\bar n}}_t}{\bar K}_t) \nonumber \\
={\kappa}^{\rm B}(T^{\rm (EM)}_{(t){\bar {\perp}}{\bar {\perp}}}+
{\hat T}_{(t)s}^{{\rm (EM)}s})+{\kappa}^{\rm S}
(T^{\rm (MA)}_{(t){\bar {\perp}}{\bar {\perp}}}+{\hat T}_{(t)s}^{{\rm (MA)}s}),
\ena
where ${\bf {\bar R}}_t$ is the Ricci tensor family and ${\bf {\bar K}}_t$ is 
the extrinsic curvature tensor family (with trace ${\bar K}_t$) of the FHSs. 
Moreover ${\kappa}^{\rm B}{\equiv}8{\pi}G^{\rm B}/c^4$, 
${\kappa}^{\rm S}{\equiv}8{\pi}G^{\rm S}/c^4$, a ``hat'' denotes an object 
projected into the FHSs and the symbol `${\mid}$' denotes a space covariant 
derivative (compatible with ${\bf {\bar h}}_t$). The values of $G^{\rm B}$ and 
$G^{\rm S}$ are by convention chosen as those measured in (hypothetical) local 
gravitational experiments in an empty Universe at epoch $t_0$. Note that 
${\bf {\bar a}}_{\cal F}$ is an object intrinsic to the FHSs. Also note that all
quantities correspond to the metric family ${\bf {\bar g}}_t$. 

Except for the non-universal coupling, the field equation (20) is similar to 
its counterpart among the various projections of the Einstein field equations 
in canonical GR. Now it would seem natural to postulate a second set of field 
equations, also yielding a natural correspondence with GR, by adopting those
projections of the Einstein equations involving the quantity
${\bar R}_{(t)j{\bar {\perp}}}$. That is, it would be tempting to postulate a 
coupling of ${\bar R}_{(t)j{\bar {\perp}}}$ directly to 
${\bar T}^{{\rm (EM)}}_{(t)j{\bar {\perp}}}$ 
and ${\bar T}^{{\rm (MA)}}_{(t)j{\bar {\perp}}}$. Unfortunately, this does not work
since it can be shown that this implies that a subset of the local conservation
laws and the corresponding subset of the contracted Bianchi identities (i.e., 
equations (26) and (36) below), would be inconsistent in the weak-field limit.

To arrive at somewhat similar field equations but such that no obvious 
inconsistencies appear, we shall take an alternative approach. First we define 
the vector field family ${\bf {\bar m}}_t$ by its components expressed in a 
GTCS, i.e.,
\eqa
{\bf {\bar m}}_t{\equiv}
-{\frac{1}{{\bar N}_t}}{\frac{\partial}{{\partial}x^0}}-{\frac{t_0}{t}}
{\frac{{\bar N}^i_{(t)}}{{\bar N}_t}}{\frac{\partial}{{\partial}x^i}}=
-{\bf {\bar n}}_t-2{\frac{t_0}{t}}{\frac{{\bar N}^i_{(t)}}{{\bar N}_t}}
{\frac{\partial}{{\partial}x^i}}, \nonumber \\ 
{\bar m}_{(t)}^{\nu}{\bar m}_{(t){\nu}}=
-1+4{\bar N}^i_{(t)}{\bar N}^k_{(t)}{\tilde h}_{(t)ik}, \qquad
{\bar m}_{(t)}^{\nu}{\bar n}_{(t){\nu}}=1.
\ena
Next we use equation (21) to define the space tensor family ${\bf {\bar L}}_t$
via its components in a GTCS, i.e.,
\eqa
{\bar L}_{(t)ij}{\equiv}
-{\frac{1}{2{\bar N}_t}}{\cal L}_{{\bar N}_t{\bf {\bar m}}_t}
{\bar h}_{(t)ij}={\bar K}_{(t)ij}+
{\frac{1}{{\bar N}_t}}{\frac{\partial}{{\partial}x^0}}{\bar h}_{(t)ij}, \quad
{\bar L}_t{\equiv}{\bar K}_t+{\frac{{\bar h}_{(t)}^{ik}}{{\bar N}_t}}
{\frac{\partial}{{\partial}x^0}}{\bar h}_{(t)ik}.
\ena
One may interpret ${\bf {\bar L}}_t$ as some sort of ``time-reversed''
extrinsic curvature tensor family. The wanted field equation set, having
the properties mentioned above, is then obtained by coupling matter fields to
the quantity ${\bar L}_{(t)j{\mid}s}^s-{\bar L}_t,_j$ rather than to
${\bar R}_{(t)j{\bar {\perp}}}={\bar K}_{(t)j{\mid}s}^s-{\bar K}_t,_j$, i.e.,
\eqa
{\bar R}_{(t)j{\bar {\perp}}}+{\Big (}{\frac{{\bar h}_{(t)}^{ik}}{{\bar N}_t}}
{\frac{\partial}{{\partial}x^0}}{\bar h}_{(t)ij}{\Big )}_{{\mid}k}-
{\Big (}{\frac{{\bar h}_{(t)}^{ik}}{{\bar N}_t}}
{\frac{\partial}{{\partial}x^0}}{\bar h}_{(t)ik}{\Big )},_j
={\bar L}_{(t)j{\mid}i}^i-{\bar L}_t,_j 
={\kappa}^{\rm B}T^{{\rm (EM)}}_{(t)j{\bar {\perp}}}
+{\kappa}^{\rm S}T^{\rm (MA)}_{(t)j{\bar {\perp}}}.
\ena
To get any further and find the rest of the wanted field equations, it is 
useful to identify the cases where we can set
${\bar N}^k_{(t)}=0$ and ${\tilde h}_{(t)ks}=S_{ks}$ in equation (13), where 
$S_{ks}dx^kdx^s$ is the metric of the 3-sphere (with radius equal to $ct_0$).
For these cases, ${\bf {\bar g}}_t$ is conformal to a metric family with 
geometry ${\bf S}^3{\times}{\bf R}$. The latter is {\em metrically static}, 
i.e., it is static except for the global time dependence of the spatial
geometry on $t$. The special case where ${\bar N}_t=1$ is a vacuum
solution of equation (20); in fact this solution represents an empty universe
(see section 4.1). More generally, if ${\bar N}_t$ has no time dependence, 
${\bf {\bar g}}_t$ will be metrically static as well, and the extrinsic
curvature will vanish, i.e., ${\bf {\bar K}}_t=0$. For a fully coupled 
quasi-metric theory somewhat similiar to GR, it would be natural to identify 
metrically static cases where ${\tilde h}_{(t)ks}=S_{ks}$ with vacuum solutions, 
where ${\bf {\bar g}}_t$ is determined solely from ${\bar N}_t$. Such solutions
are fully determined from the scale factor ${\bar F}_t$.

Now one naturally expects that a complete set of quasi-metric field equations
should represent a full coupling between space-time curvature and ${\bf T}_t$, 
in addition to being compatible with equation (13). That is, we would expect to 
find a new space-time tensor family ${\bf {\bar Q}}_t$ defined from its 
projections ${\bar Q}_{(t){\bar {\perp}}{\bar {\perp}}}$,
${\bar Q}_{(t)j{\bar {\perp}}}={\bar L}_{(t)j{\mid}i}^i-{\bar L}_t,_j$ and
${\bar Q}_{(t)ij}$ with respect to the FHSs. These projections are expected to 
play almost the same role as do the projections of the Einstein tensor in 
canonical GR. However, it is important to notice that unlike 
${\bf {\bar R}}_t$ and the Einstein tensor family ${\bf {\bar G}}_t$, the 
definition of ${\bf {\bar Q}}_t$ depends directly on the geometry of the FHSs 
and their extrinsic curvature. This means that the expected expressions for the
projections of ${\bf {\bar Q}}_t$ will not be exactly valid for any 
hypersurfaces other than the FHSs. In contrast, in canonical GR, the 
projections of the Einstein tensor ${\bf {\bar G}}$ on a Lorentzian manifold 
with metric ${\bf {\bar g}}$ take the same general form valid for any foliation
of ${\bf {\bar g}}$ into spatial hypersurfaces. The projections of 
${\bf {\bar G}}_t$ with respect to the FHSs also take this general form (see 
e.g., [2] and references therein)
\eqa
{\bar G}_{(t){\bar {\perp}}{\bar {\perp}}}={\frac{1}{2}}({\bar P}_t
+{\bar K}_t^2-{\bar K}_{(t)ks}{\bar K}_{(t)}^{ks}), \qquad
{\bar G}_{(t){\bar {\perp}}j}{\equiv}{\bar R}_{(t){\bar {\perp}}j}
={\bar K}_{(t)j{\mid}i}^i-{\bar K}_t,_j,
\ena
\eqa
{\bar G}_{(t)ij}=-{\frac{1}{{\bar N}_t}}
{\cal L}_{{\bar N}_t{\bf {\bar n}}_t}({\bar K}_{(t)ij}-
{\bar K}_t{\bar h}_{(t)ij})+3{\bar K}_t{\bar K}_{(t)ij}-
{\frac{1}{2}}({\bar K}_t^2+
{\bar K}_{(t)ks}{\bar K}_{(t)}^{ks}){\bar h}_{(t)ij} \nonumber \\
-2{\bar K}_{(t)is}{\bar K}_{(t)j}^s
-c^{-2}{\bar a}_{{\cal F}i{\mid}j}-c^{-4}{\bar a}_{{\cal F}i}
{\bar a}_{{\cal F}j}+(c^{-2}{\bar a}^s_{{\cal F}{\mid}s}
+c^{-4}{\bar a}_{\cal F}^s{\bar a}_{{\cal F}s}){\bar h}_{(t)ij}+
{\bar H}_{(t)ij},
\ena
where ${\bf {\bar H}}_t$ is the spatial Einstein tensor family intrinsic to 
the FHSs.

We will now require that ${\bf {\bar Q}}_t$ and ${\bf {\bar G}}_t$ should have 
somewhat similar dynamical structures. That is, ${\bar Q}_{(t)ij}$ and 
${\bar G}_{(t)ij}$ should both have common (up to signs) second order terms 
$-{\frac{1}{{\bar N}_t}}{\cal L}_{{\bar N}_t{\bf {\bar n}}_t}{\bar K}_{(t)ij}$ and 
${\bar H}_{(t)ij}$ in equation (25). Furthermore, we must have that 
${\bar Q}_{(t){\bar {\perp}}{\bar {\perp}}}+
{\hat{\bar Q}}^s_{(t)s}=2{\bar R}_{(t){\bar {\perp}}{\bar {\perp}}}$ to fulfil equation
(20), and ${\bar Q}_{(t){\bar {\perp}}j}={\bar L}_{(t)j{\mid}i}^i-{\bar L}_t,_j$ to 
fulfil equation (23). Besides, for the metrically static vacuum cases, the 
equation ${\bar Q}_{(t)ij}=0$ should yield the relationship ${\bar H}_{(t)ij}+
c^{-2}{\bar a}_{{\cal F}i{\mid}j}+c^{-4}{\bar a}_{{\cal F}i}{\bar a}_{{\cal F}j}-
(c^{-2}{\bar a}_{{\cal F}{\mid}s}^s-{\frac{1}{(ct{\bar N}_t)^2}}){\bar h}_{(t)ij}=0$,
which follows directly from equation (13) for these cases. But the extrinsic 
curvature also vanishes identically for metrically static interiors (where we 
expect ${\tilde h}_{(t)ks}{\neq}S_{ks}$), so this means that we should have 
${\bar Q}_{(t)ij}=-c^{-2}{\bar a}_{{\cal F}i{\mid}j}-
c^{-4}{\bar a}_{{\cal F}i}{\bar a}_{{\cal F}j}+(c^{-2}{\bar a}_{{\cal F}{\mid}s}^s
-{\frac{1}{(ct{\bar N}_t)^2}}){\bar h}_{(t)ij}-{\bar H}_{(t)ij}$ and thus
${\bar Q}_{(t){\bar {\perp}}{\bar {\perp}}}=-{\frac{1}{2}}{\bar P}_t
+3c^{-4}{\bar a}_{{\cal F}s}{\bar a}_{\cal F}^s+{\frac{3}{(ct{\bar N}_t)^2}}$
for the metrically static cases. (The other sign for ${\bar Q}_{(t)ij}$ cannot 
be chosen since we for physical reasons should have that in general,
${\bar Q}_{(t){\bar {\perp}}{\bar {\perp}}}>0$ and ${\hat{\bar Q}}^s_{(t)s}>0$ for 
metrically static interiors. That is, we expect these quantities to be 
non-negative since they should be coupled to suitable projections of 
${\bf T}_t$.)

However, at this point a crucial problem arises due to the contracted Bianchi
identities ${\bar G}_{(t){\mu};{\nu}}^{\nu}{\equiv}0$ (where a semicolon denotes 
taking a metric covariant derivative in component notation, holding $t$ fixed). 
Projected with respect to the FHSs, these identities read (see, e.g., [2])
\eqa
{\cal L}_{{\bf {\bar n}}_t}{\bar G}_{(t){\bar {\perp}}{\bar {\perp}}}=
{\bar K}_t{\bar G}_{(t){\bar {\perp}}{\bar {\perp}}}+{\bar K}_{(t)}^{ks}
{\bar G}_{(t)ks}-2c^{-2}{\bar a}_{\cal F}^s{\bar G}_{(t){\bar {\perp}}s}
-{\hat {\bar G}}^s_{(t){\bar {\perp}}{\mid}s},
\ena
\eqa
{\frac{1}{{\bar N}_t}}{\cal L}_{{\bar N}_t{\bf {\bar n}}_t}
{\bar G}_{(t)j{\bar {\perp}}}={\bar K}_t{\bar G}_{(t)j{\bar {\perp}}}
-c^{-2}{\bar a}_{{\cal F}j}{\bar G}_{(t){\bar {\perp}}{\bar {\perp}}}
-c^{-2}{\bar a}_{\cal F}^s{\bar G}_{(t)js}-{\hat {\bar G}}^s_{(t)j{\mid}s}.
\ena
That is, it turns out that equations (23) and (27), in combination with the 
deduced expressions for ${\bar Q}_{(t){\bar {\perp}}{\bar {\perp}}}$ and 
${\bar Q}_{(t)ij}$ for metrically static interiors, yield the wrong Newtonian 
limit, so that equation (27) does not correspond with its counterpart Euler 
equation (see equation (37) below). In fact, the only way to avoid this problem
while still keeping the relationship ${\bar Q}_{(t){\bar {\perp}}{\bar {\perp}}}+
{\hat{\bar Q}}^s_{(t)s}=2{\bar R}_{(t){\bar {\perp}}{\bar {\perp}}}$, is to choose
${\bar Q}_{(t){\bar {\perp}}{\bar {\perp}}}=2{\bar R}_{(t){\bar {\perp}}{\bar {\perp}}}$. 
This choice comes with the extra condition ${\hat {\bar Q}}^s_{(t)j{\mid}s}-c^{-2}
{\bar a}_{\cal F}^s{\bar Q}_{(t)js}=0$ obtained from equation (27). However, this 
still yields no possible consistent coupling of $T_{(t)ij}$ to ${\bar Q}_{(t)ij}$ 
given equation (37) below, so to avoid said problem we are forced to set 
${\bar Q}_{(t)ij}=0$. Thus there can be no extra scalar field equation besides 
equation (20) and also no additional spatial tensor equation representing a
full coupling to the spatial projections of ${\bf T}_t$. In other words, we 
have found that {\em it is not possible to construct a viable, fully coupled 
quasi-metric gravitational theory.} 

Nevertheless, fortunately it is still possible to have a manifestly traceless 
field equation ${\bar Q}_{(t)ij}=0$ having the desired dynamical properties, in 
addition to being compatible with equation (27) (and where the couplig to 
${\bf T}_t$ is via equation (20)). The choice of terms quadratic in extrinsic 
curvature in such an equation would seem somewhat uncertain, but this question 
can be resolved by a restriction involving a particular projection of the Weyl 
tensor family ${\bf {\bar C}}_t$. That is, we require that the projection
${\bar C}_{(t){\bar {\perp}}i{\bar {\perp}}j}$ should be determined from the intrinsic
geometry of the FHSs alone, with no explicit dependence on extrinsic curvature
(or on ${\bf {\bar a}}_{\cal F}$). Thus we define a (unique) relationship having
this property, i.e.,
\eqa
{\bar C}_{(t){\bar {\perp}}i{\bar {\perp}}j}={\tilde H}_{(t)ij}+
{\frac{1}{(ct{\bar N}_t)^2}}{\bar h}_{(t)ij},
\ena
where ${\bf {\tilde H}}_t$ is the spatial Einstein tensor family calculated 
from the metric family ${\bf {\tilde h}}_t$. (Note that the 
foliation-dependence of equation (28) (and thus of the field equations) is 
directly given from its right hand side and that it does not involve extrinsic 
curvature.) Moreover, we also have in general that ${\bf {\bar C}}_t$ can be 
expressed by the Riemann tensor family, the Ricci tensor family and the Ricci 
scalar family ${\bar R}_t$. In particular, this yields 
\eqa
{\bar C}_{(t){\bar {\perp}}i{\bar {\perp}}j}=
{\bar R}_{(t){\bar {\perp}}i{\bar {\perp}}j}+{\frac{1}{2}}{\bar R}_{(t)ij}
-{\frac{1}{2}}{\Big (}{\bar R}_{(t){\bar {\perp}}{\bar {\perp}}}+{\frac{1}{3}}
{\bar R}_t{\Big )}{\bar h}_{(t)ij}, \nonumber \\
{\bar R}_{(t){\bar {\perp}}i{\bar {\perp}}j}={\frac{1}{{\bar N}_t}}
{\cal L}_{{\bar N}_t{\bf {\bar n}}_t}{\bar K}_{(t)ij}+
{\bar K}_{(t)i}^{s}{\bar K}_{(t)sj}+c^{-2}{\bar a}_{{\cal F}i{\mid}j}+
c^{-4}{\bar a}_{{\cal F}i}{\bar a}_{{\cal F}j}, \nonumber \\
{\bar R}_t={\bar P}_t-2{\cal L}_{{\bf {\bar n}}_t}{\bar K}_t
+{\bar K}_{(t)ks}{\bar K}_{(t)}^{ks}+{\bar K}_t^2
-2c^{-2}{\bar a}_{{\cal F}{\mid}s}^s
-2c^{-4}{\bar a}_{{\cal F}}^s{\bar a}_{{\cal F}s}.
\ena
Equation (29) may now be inserted into equation (28) to give a definition of 
${\bar Q}_{(t)ij}$ via the quantities ${\bar G}_{(t)ij}$, 
${\bar R}_{(t){\bar {\perp}}{\bar {\perp}}}$ and 
${\bar G}_{(t){\bar {\perp}}{\bar {\perp}}}$. That is, we define ${\bar Q}_{(t)ij}$ 
from equation (25) and the requirement that
\eqa
{\bar G}_{(t)ij}=-{\bar Q}_{(t)ij}-2c^{-2}{\bar a}_{{\cal F}i{\mid}j}-
2c^{-4}{\bar a}_{{\cal F}i}{\bar a}_{{\cal F}j}
-2{\bar K}_{(t)i}^{s}{\bar K}_{(t)sj}+2{\bar K}_t{\bar K}_{(t)ij}
\nonumber \\
+{\frac{1}{3}}{\Big [}2{\bar R}_{(t){\bar {\perp}}{\bar {\perp}}}
-{\bar G}_{(t){\bar {\perp}}{\bar {\perp}}}+2c^{-2}{\bar a}_{{\cal F}{\mid}s}^s
+2c^{-4}{\bar a}_{{\cal F}}^s{\bar a}_{{\cal F}s}
+2{\bar K}_{(t)ks}{\bar K}_{(t)}^{ks}-2{\bar K}_t^2
{\Big ]}{\bar h}_{(t)ij}.
\ena
We then get the definition
\eqa
{\bar Q}_{(t)ij}{\equiv}{\frac{1}{{\bar N}_t}}{\cal L}_{{\bar N}_t
{\bf {\bar n}}_t}{\bar K}_{(t)ij}+
{\frac{1}{3}}{\Big [}2{\bar K}_{(t)ks}{\bar K}_{(t)}^{ks}-{\bar K}_t^2
-{\cal L}_{{\bf {\bar n}}_t}{\bar K}_t{\Big ]}{\bar h}_{(t)ij}
\nonumber \\
+{\bar K}_t{\bar K}_{(t)ij}-c^{-2}{\bar a}_{{\cal F}i{\mid}j}-
c^{-4}{\bar a}_{{\cal F}i}{\bar a}_{{\cal F}j}
+{\Big [}c^{-2}{\bar a}_{{\cal F}{\mid}s}^s
-{\frac{1}{(ct{\bar N}_t)^2}}{\Big ]}{\bar h}_{(t)ij}-{\bar H}_{(t)ij}=0,
\ena
where the requirement on the spatial Ricci curvature scalar family 
${\bar P}_t$,
\eqa
{\bar P}_t=-4c^{-2}{\bar a}_{{\cal F}{\mid}s}^s
+2c^{-4}{\bar a}_{{\cal F}}^s{\bar a}_{{\cal F}s}+{\frac{6}{(ct{\bar N}_t)^2}},
\ena
ensures that equation (31) is indeed manifestly traceless. Besides, the 
components of the spatial Einstein tensor family ${\bf {\bar H}}_t$ are given 
by
\eqa
{\bar H}_{(t)ij}=-c^{-2}{\bar a}_{{\cal F}i{\mid}j}-
c^{-4}{\bar a}_{{\cal F}i}{\bar a}_{{\cal F}j}+
c^{-2}{\bar a}_{{\cal F}{\mid}s}^s{\bar h}_{(t)ij}+{\tilde H}_{(t)ij}.
\ena
Note that, while equation (32) implies that 
${\tilde P}_t={\frac{6}{(ct_0)^2}}$, we have that ${\tilde H}_{(t)ij}$ is not 
necessarily equal to $-{\frac{1}{(ct_0)^2}}{\tilde h}_{(t)ij}$. This shows that, 
while there is prior 3-geometry, there is still some dynamical freedom 
associated with the ``basic'' line element family $d{\tilde{\sigma}}_t^2$. This
is further illustrated by writing equation (31) in the form (using equations 
(20) and (33))
\eqa
{\frac{1}{{\bar N}_t}}{\cal L}_{{\bar N}_t{\bf {\bar n}}_t}{\bar K}_{(t)ij}
+{\bar K}_t{\bar K}_{(t)ij}-{\tilde H}_{(t)ij} \nonumber \\
={\frac{1}{3}}{\Big [}{\bar R}_{(t){\bar {\perp}}{\bar {\perp}}}
+{\bar K}_t^2-{\bar K}_{(t)ks}{\bar K}_{(t)}^{ks}
-c^{-2}{\bar a}_{{\cal F}{\mid}s}^s
-c^{-4}{\bar a}_{{\cal F}}^s{\bar a}_{{\cal F}s}+{\frac{3}{(ct{\bar N}_t)^2}}
{\Big ]}{\bar h}_{(t)ij}.
\ena
We notice that taking the trace of equation (34) recovers the (general)
expression (20) for ${\bar R}_{(t){\bar {\perp}}{\bar {\perp}}}$.

Equations (20), (23) and (31) determine ${\bar Q}_{(t){\bar {\perp}}{\bar {\perp}}}
{\equiv}2{\bar R}_{(t){\bar {\perp}}{\bar {\perp}}}$, ${\bar Q}_{(t){\bar {\perp}}j}
{\equiv}{\bar L}_{(t)j{\mid}i}^i-{\bar L}_t,_j$ and 
${\bar Q}_{(t)ij}$, respectively. This determines 9 of the 20 independent 
components of the Riemann tensor family in 4 dimensions. Equation (32) yields
an extra restriction so that all together, said equations determine 10 of said
components, the same number as the for the full Einstein tensor in ordinary 
GR. Besides, for QMR we see from equation (28) that 5 of the 10 independent 
components of the Weyl tensor family are determined in addition to 5 of the 
10 independent components of the Ricci tensor family. On the other hand, in GR
the field equations determine the Ricci tensor in full, leaving the Weyl tensor 
free.

The full set of quasi-metric field equations then consists of equations (20),
(23), (32) and (34). (Equation (28) or (31) can alternatively be substituted 
for equation (34).) Note that these field equations are valid only for 
projections with respect to the FHSs; they do not hold exactly for projections 
with respect to any other hypersurfaces (for a further discussion of this, see 
the end of this section). Also note that the quasi-metric field equations have 
a somewhat similar split-up as Einstein's field equations into dynamical 
equations and constraints. That is, equations (23) and (32) represent 4 
constraint equations while equations (20) and (34) represent 6 dynamical 
equations, the same numbers as for GR. However, the Einstein equations include 
no counterpart to equation (32) but rather an extra scalar constraint 
corresponding to equation (24); such an equation is missing in quasi-metric 
gravity.

Similar to Einstein's equations, the dynamical gravitational field may be 
taken to be the spatial metric family ${\bf {\bar h}}_t$, representing two 
independent propagating dynamical degrees of freedom. However, according to 
equation (13) it is possible to split up ${\bf {\bar h}}_t$ further into two 
separate dynamical fields, i.e., one dynamical scalar field represented by the 
lapse function family ${\bar N}_t$, plus ${\bf {\tilde h}}_t$, which is a 
dynamical tensor field. The propagating dynamical degrees of freedom should be 
associated with the trace-free part of the latter. Finally, we notice that all 
field equations are time-reversal invariant.

The field equations make it possible to calculate ${\bf {\bar g}}_t$ from the 
projections of the physical source ${\bf T}_t$ with respect to the FHSs. And as
mentioned above, said equations are in principle valid only for the FHSs as 
long as the global time function $t$ is unique. However, the uniqueness of $t$ 
follows from the topological structure of quasi-metric space-time, since the 
FHSs are defined to be compact with positive curvature scalar ${\tilde P}_t$.  
In quasi-metric cosmology, this singles out the cosmic rest frame (the frame 
where the cosmic relic microwave radiation is measured to be isotropic on 
average) as a natural ``preferred frame'' since the FOs should be at rest on 
average with respect to this frame. Thus when doing cosmology, the global time 
function and the FHSs are given {\em a priori} from the postulated form of 
quasi-metric space-time.

But for local, isolated systems, applications of the field equations would seem
to be limited in practice since they are expressed in terms of one particular 
foliation of quasi-metric space-time into spatial hypersurfaces, apparently 
involving the cosmic rest frame. Therefore, for isolated systems, one may
substitute the condition ${\tilde P}_t={\frac{6}{(ct_0)^2}}$ with the 
approximate alternative condition ${\tilde P}_t=0$. This means that the FHSs 
may be taken to be approximately flat sufficienly far from an isolated system.
But if the FHSs are taken to be asymptotically flat, this means that the global
time function will no longer be unique. In fact, it will then be possible to 
define an alternative global time function $t'={x^0}^{'}/c$ and an alternative 
foliation of ${\bf {\bar g}}_{t^{'}}$ into an alternative set of spatial 
hypersurfaces (also being asymptotically flat). An alternative class of 
observers always moving orthogonally to the alternative hypersurfaces may then 
be defined such that said observers are at rest with respect to the barycentre 
of the isolated system. Moreover, the field equations (with 
${\tilde P}_{t'}=0$) may then be transformed with respect to this new set of 
hypersurfaces. However, the field equations would not be invariant under said 
transformation; they would depend on the velocity of the isolated system 
with respect to the cosmic rest frame. In practice the 
``preferred frame''-effects introduced by said procedure should be small (at 
most of post-Newtonian order), if the size of the isolated system is small 
compared to $ct_0$ and its local speed with respect to the cosmic rest frame is
much smaller than the speed of light.
\subsection{Local conservation laws}
Within the metric framework, one usually just substitutes partial derivatives 
with covariant derivatives when generalizing differential laws from flat to 
curved space-time. In fact, this rule in the form ``comma goes to semicolon'' 
follows directly from the EEP in most metric theories [1]. But in 
quasi-metric theory it is possible to couple non-gravitational fields to first
derivatives of the scale factor of the FHSs such that the EEP still holds. 
That is, any coupling of non-gravitational fields to the fields 
${\bar a}_{{\cal F}j}$, ${\frac{{\bar N}_t,_{{\bar {\perp}}}}{{\bar N}_t}}$,
${\frac{{\bar N}_t,_t}{{\bar N}_t}}$ and ${\frac{1}{t}}$ may be made to 
vanish in the local inertial frames so that these couplings do not
interfere with the local non-gravitational physics.

In particular, the EEP implies that the local conservation laws take the form 
${\bf {\nabla}}{\bf {\cdot}}{\bf T}=0$ in any metric theory based on an
invariant action principle, independent of the field equations [1]. The 
reason why the conservation laws must take this form is that they then imply 
that inertial test particles move on geodesics of the metric. So, in said 
metric theories, the above form of the local conservation laws is sufficient to
ensure that they are consistent with the equations of motion. But in 
quasi-metric theory, consistency with the equations of motion does not 
necessarily imply that the local conservation laws take the form shown above. 
This fact, in addition to the possibility of extra couplings between 
non-gravitational fields and the fields ${\bar a}_{{\cal F}j}$, 
${\frac{{\bar N}_t,_{{\bar {\perp}}}}{{\bar N}_t}}$,
${\frac{{\bar N}_t,_t}{{\bar N}_t}}$ and ${\frac{1}{t}}$, means that the EEP 
does not necessarily imply a form similar to ${\nabla}{\bf {\cdot}}{\bf T}=0$ 
of the local conservation laws in quasi-metric theory. That is, the 
divergence ${\stackrel{{\hbox{\tiny$\star$}}}{\bf {\bar {\nabla}}}}
{\bf {\cdot}}{\bf T}_t$ will in general not vanish (nor will
${\bf {\bar {\nabla}}}{\bf {\cdot}}{\bf T}_t$ with $t$ fixed), so the EEP is 
insufficient to determine the form of the local conservation laws in QMR.

Since the EEP is not sufficient to determine the form of the local 
conservation laws in quasi-metric theory, we have to deduce their form from 
other criteria. That is, in order to have the correct Newtonian limit in 
addition to being consistent with electromagnetism coupled to gravity [4], 
the local conservation laws (with $t$ fixed) must take the form
\eqa
T_{(t){\mu};{\nu}}^{\nu}=
2{\frac{{\bar N}_t,_{\nu}}{{\bar N}_t}}T_{(t){\mu}}^{\nu}
=2c^{-2}{\bar a}_{{\cal F}s}{\hat T}^s_{(t){\mu}}
-2{\frac{{\bar N}_t,_{\bar {\perp}}}{{\bar N}_t}}T_{(t){\bar {\perp}}{\mu}}.
\ena
By projecting equation (35) with respect to the FHSs we get (see, e.g., [2] 
and references therein for general projection formulae)
\eqa
{\cal L}_{{\bf {\bar n}}_t}T_{(t){\bar {\perp}}{\bar {\perp}}}=
{\Big (}{\bar K}_t-2{\frac{{\bar N}_t,_{\bar {\perp}}}{{\bar N}_t}}
{\Big )}T_{(t){\bar {\perp}}{\bar {\perp}}}
+{\bar K}_{(t)ks}{\hat T}_{(t)}^{ks}-{\hat T}^s_{(t){\bar {\perp}}{\mid}s},
\ena
\eqa
{\frac{1}{{\bar N}_t}}{\cal L}_{{\bar N}_t{\bf {\bar n}}_t}
T_{(t)j{\bar {\perp}}}={\Big (}{\bar K}_t
-2{\frac{{\bar N}_t,_{\bar {\perp}}}{{\bar N}_t}}
{\Big )}T_{(t)j{\bar {\perp}}}-c^{-2}{\bar a}_{{\cal F}j}
T_{(t){\bar {\perp}}{\bar {\perp}}}
+c^{-2}{\bar a}_{{\cal F}s}{\hat T}_{(t)j}^s-{\hat T}^s_{(t)j{\mid}s}.
\ena
Moreover, for the non-metric part of the connection, i.e., as a counterpart to 
equation (35), an extra local ``conservation law'' can be found by calculating 
the quantity $c^{-1}{\stackrel{{\hbox{\tiny$\star$}}}{\bf {\bar {\nabla}}}}
_{{\!}{\!}{\frac{\partial}{{\partial}t}}}T_{(t){\mu}}^0$. Assuming that the 
$t$-dependence of ${\bf T}_t$ can be found from the special case of an 
expanding source consisting of noninteracting particles and fields (see 
section 3.2), we find that
\eqa
T_{(t){\mu}{\bar *}t}^0=-{\frac{2}{{\bar N}_t}}{\Big (}{\frac{1}{t}}+
{\frac{{\bar N}_t,_t}{{\bar N}_t}}{\Big )}T_{(t){\bar {\perp}}{\mu}},
\ena
where the symbol `${\bar *}$' denotes a degenerate covariant derivative 
compatible with the metric family ${\bf {\bar g}}_t$. By applying equation 
(35) to a source consisting of a perfect fluid with no pressure 
(i.e., dust), and projecting the resulting equations with the quantity
${\bf {\bar g}}_t+c^{-2}{\bf {\bar u}}_t{\otimes}{\bf {\bar u}}_t$, we find 
that the dust particles move on geodesics of 
${\stackrel{{\hbox{\tiny$\star$}}}{\bf {\bar {\nabla}}}}$ in 
$({\cal N},{\bf {\bar g}}_t)$. (It is sufficient to use equation (35) since we
have the counterpart in $({\cal N},{\bf {\bar g}}_t)$ to equation (8) with
${\stackrel{{\hbox{\tiny$\star$}}}{\bf {\bar {\nabla}}}}
_{{\!}{\!}{\frac{\partial}{{\partial}t}}}{\bf {\bar u}}_t=0$.) This guarantees that the 
dust particles move on geodesics of 
${\stackrel{{\hbox{\tiny$\star$}}}{\bf{\nabla}}}$ in $({\cal N},{\bf g}_t)$ as 
well; see section 3.6 for justification. Besides, since ${\bf T}_t$ is the 
active stress-energy tensor and since no extra field independent of 
${\bf {\bar g}}_t$ couples to gravitating bodies in QMR, this result should 
apply even to (sufficiently small) dust particles with significant 
self-gravitational energy and not only to test particles. This means that, 
although $G^{\rm B}_t$ and $G^{\rm S}_t$ are formally variable in QMR, 
no Nordtvedt effect should be associated with this formal variability.

Since ${\bf T}_t$ is not directly measurable locally one must know how it 
relates to the {\em passive stress-energy tensor} ${\cal T}_t$ in 
$({\cal N},{\bf g}_t)$, or equivalently, to the passive stress-energy tensor 
${\bf {\bar {\cal T}}}_t$ in $({\cal N},{\bf {\bar g}}_t)$ (which can be 
measured locally using atomic units). This means that equations (35)-(38) 
do not represent the ``more physical'' local conservation laws involving 
${\cal T}_t$. Besides, the local conservation laws shown in (35)-(38) are 
compatible with ${\bf {\bar g}}_t$ and not with ${\bf g}_t$. However, said 
more physical local conservation laws can be found by calculating 
${\stackrel{{\hbox{\tiny$\star$}}}{\nabla}}{\bf {\cdot}}{\cal T}_t$ when 
${\bf g}_t$ is known. But these more physical local conservation laws take no 
predetermined form.

The relationship between ${\bf T}_t$ and ${\cal T}_t$ (or 
${\bf {\bar {\cal T}}}_t$) depends in principle explicitly on the general 
nature of the matter source. For example, this relationship will be different 
for a perfect fluid consisting of material particles than for electromagnetic 
radiation. To illustrate this we may consider ${\bf T}_t$ for a perfect fluid:
\eqa
{\bf T}_t=({\tilde {\varrho}}_{\rm m}+
c^{-2}{\tilde {p}}){\bf {\bar u}}_t{\otimes}
{\bf {\bar u}}_t + {\tilde p}{\bf {\bar g}}_t, 
\ena
where ${\tilde {\varrho}}_{\rm m}$ is the active mass-energy density in the 
local rest frame of the fluid and ${\tilde p}$ is the active pressure. The 
corresponding expressions for ${\cal T}_t$ and ${\bf {\bar {\cal T}}}_t$ are
\eqa
{\cal T}_t={\sqrt{\frac{{\bar h}_t}{h_t}}}{\Big [}
({\varrho}_{\rm m}+c^{-2}{p}){\bf u}_t{\otimes}{\bf u}_t+p{\bf g}_t{\Big ]}, 
\qquad {\bf {\bar {\cal T}}}_t=({\varrho}_{\rm m}+
c^{-2}p){\bf {\bar u}}_t{\otimes}{\bf {\bar u}}_t + p{\bf {\bar g}}_t,
\ena
where ${\varrho}_{\rm m}$ is the passive mass-energy density as measured in the 
local rest frame of the fluid and $p$ is the passive pressure. Also, by 
definition ${\bar h}_t$ and $h_t$ are the determinants of ${\bf {\bar h}}_t$ 
and ${\bf h}_t$, respectively. Now the relationship between 
${\tilde {\varrho}}_{\rm m}$ and ${\varrho}_{\rm m}$ is given by
\eqa
{\varrho}_{\rm m}=
\left\{
\begin{array}{ll}
{\frac{t_0}{t}}{\bar N}_t^{-1}{\tilde {\varrho}}_{\rm m}, 
& \qquad $$\hbox{for a fluid of material particles,}$$ \\[1.5ex]
{\frac{t_0^2}{t^2}}{\bar N}_t^{-2}{\tilde {\varrho}}_{\rm m}, 
& \qquad $$\hbox{for electromagnetic fields,}$$
\end{array}
\right.
\ena
and a similar relationship exists between ${\tilde p}$ and $p$. The reason why
the relationship between ${\tilde {\varrho}}_{\rm m}$ and ${\varrho}_{\rm m}$ is 
different for the electromagnetic field than for material fluid sources, is 
that the cosmological redshift directly influences passive electromagnetic
mass-energy but not the passive mass-energy of a material fluid.
\subsection{The quasi-metric initial-value problem}
Equation sets (24)-(25) in combination with the Einstein field equations,
respectively equations (20), (23) plus (31) (or one of (28) and (34)), 
considered as dynamical systems for initial value problems, share some 
similiarities but are also dissimilar in crucial ways. Both equation sets 
consist of dynamical equations plus constraints. The constraint equations are 
determined by initial data on an initial FHS (or, for the set (24)-(25), 
initial data on an arbitrary spatial hypersurface), whereas the dynamical 
equations are not. Now we see that the quantities 
${\bar G}_{(t){\bar {\perp}}{\bar {\perp}}}$, ${\bar L}_{(t)ij}$ and 
${\bar R}_{(t){\bar {\perp}}j}{\equiv}{\bar G}_{(t){\bar {\perp}}j}$ are all determined
by the initial data, while the quantities 
${\bar R}_{(t){\bar {\perp}}{\bar {\perp}}}$, ${\bar G}_{(t)ij}$ and ${\bar Q}_{(t)ij}$ 
are not. Thus for the quasi-metric system, equations (23) and (32) represent 
the constraints whereas equations (20) and (31) (or one of (28) and 
(34)) represent the dynamical equations. This means that both initial value
systems contain 4 constraint equations and 6 dynamical equations. However,
as opposed to its counterpart valid for the dynamical system (24)-(25), out
of equations (20) and (34), only equation (20) represents a dynamically 
independent coupling to matter sources. 
That is, a crucial difference between said dynamical systems is that 
equation (34) is only partially coupled to matter sources via the scalar 
quantity ${\bar R}_{(t){\bar {\perp}}{\bar {\perp}}}$ and equation (20). On the other
hand, for GR, equation (25) is fully coupled to the spatial projection 
$T_{(t)ij}$ of the stress-energy tensor via the Einstein field equations. (One 
consequence of this difference is that quasi-metric interior solutions will be 
less dependent on the source's equation of state than for comparable situations
in GR, so that any quasi-metric interior solution should cover a wider range 
of physical conditions than its counterparts in GR.) Also there is no 
quasi-metric counterpart to the Einstein scalar constraint obtained from 
equation (24) since in quasi-metric gravity, there is no separate independent
coupling of matter fields to the quantity 
${\bar G}_{(t){\bar {\perp}}{\bar {\perp}}}$.

Now we notice that equation (26) must be consistent with equation (36) and
equation (27) must be consistent with equation (37) for every step of the
quasi-metric initial-value problem. (For weak fields, one may readily show that
these requirements are satisfied to lowest order.) Besides, we notice that, 
when specifying initial data ${\bf {\bar h}}_t$ and ${\bf {\bar K}}_t$ on an 
initial FHS, due to equation (32) there is no freedom to choose the lapse 
function family ${\bar N}_t$ independently. But then, since the FOs are moving
orthogonally with respect to the FHSs, there can be no freedom to choose the 
components ${\frac{t_0}{t}}{\bar N}_{(t)}^j$ of the shift vector family 
independently either. But this means, unlike the GR case, that the quasi-metric
initial-value system describes the time evolution of a fixed sequence of 
spatial hypersurfaces, i.e., the FHSs. That is, there is no freedom to 
choose lapse and shift as for the GR case, where the evolution of an initial 
spatial hypersurface into some fixed final one may be done by foliating 
space-time in many different ways. This means that equations (36) and (37)
are required to hold, independently of the field equations, at every subsequent
FHS of the quasi-metric initial-value problem. Thus, it cannot be expected that
the quasi-metric dynamical equations will automatically preserve the 
constraints, since there is no remaining gauge freedom to choose
the shift vector family freely. On the other hand, in GR, the vacuum Einstein 
dynamical equations preserve the constraints since the contracted Bianchi 
identities assure that there are no extra restrictions.

Also, we notice that the quantities ${\bar N}_t,_t$ and ${\tilde h}_{(t)ij},_t$ 
play no dynamical role in the quasi-metric initial-value problem since in 
principle, they can be chosen freely on an initial FHS, yet their values at 
subsequent FHSs cannot be determined from dynamical equations. Rather, to 
control the evolution of ${\bar N}_t$ and ${\tilde h}_{(t)ij}$, the values of 
said quantities must be determined independently from the effects of the 
cosmic expansion on the matter source for each time step. Finally, we notice 
that what has been described in this section, is the initial value problem in 
$({\cal N},{\bf {\bar g}}_t)$. In order to have the full initial value problem 
in $({\cal N},{\bf g}_t)$ as well, the transformation 
${\bf {\bar g}}_t{\rightarrow}{\bf g}_t$ (see the next section) must be 
performed at each time step so that equation (10) can be used to propagate the 
sources.
\subsection{Constructing ${\bf g}_t$ from ${\bf {\bar g}}_t$}
The global part of the NKE is realized explicitly in the evolution of
${\bar F}_t$, as can be seen from equation (17). But from equation (18) we see 
that space derivatives of ${\bar N}_t$ yield local contributions to the NKE 
of the FHSs as well. {\em These local contributions are not realized explicitly
in the evolution of ${\bar F}_t$, so whenever ${\bar y}_t{\neq}0$ in equation 
(18), it is necessary to construct a new metric family ${\bf g}_t$.} In the 
following we will see the reason why.

The question now is just how the metric family (13) should be modified to
include the local effects of the NKE. This question can be answered by 
noticing that according to equation (18), the local effects of the NKE 
should take the form of an ``expansion'' that varies from place to place. 
That is, the tangent spaces of the FHSs should experience a varying degree of 
expansion as a consequence of the local contribution ${\bar y}_t$ to 
${\bar H}_t$. Two points now are, that the local contribution ${\bar y}_t$
to the expansion is due to gravitation, and that this contribution is not 
reflected explicitly in the evolution of the scale factor ${\bar F}_t$ as can 
be seen from equation (17). Thus, whenever ${\bar y}_t$ is nonzero we may 
think of a change of distances in any tangent space of the FHSs as consisting 
of an expansion plus a contraction. That is, the FOs seem to ``move'' more than 
the explicit change of ${\bar F}_t$ would indicate. The modification of the 
metric family (13) then consists of a compensation for this extra 
gravitationally induced ``motion''. 

To have a consistent transformation ${\bf {\bar g}}_t{\rightarrow}{\bf g}_t$,
we need to treat the effects of the extra gravitationally induced ``motion'' 
in each tangent space, i.e., locally. To do that, it turns out that we need to 
define a 3-vector field ${\bf {\bar x}}_{\cal F}$ representing the coordinate 
distance to a local fictitious ``centre of gravity'' in each tangent space. 
This is necessary to be able to define a family of 3-vector fields ${\bf v}_t$ 
telling how much the local FO in each tangent space ``recedes'' from the local 
``centre of gravity'' due to said gravitationally induced expansion. Besides, 
since the coordinate positions of all FOs must be unaffected, the FOs must 
simultaneously ``fall'' with velocity $-{\bf v}_t$ toward the local ``centre 
of gravity'' to cancel out said ``recession''. The extra ``motion'' 
involved induces corrections in the coordinate length and time intervals as 
perceived by any FO. That is, the metric components of equation (13) in a GTCS 
must be modified to yield a new metric family ${\bf g}_t$.

In the special case where ${\bf {\bar g}}_t$ is spherically symmetric with
respect to one distinguished point, the spatial coordinates of this point 
represent a natural local ``centre of gravity'' in each tangent space of the
FHSs. In this case we obviously have ${\bf {\bar x}}_{\cal F}=r{\frac{\partial}
{{\partial}r}}$ expressed in a spherical polar GTCS where the distinguished
point lies at the origin. It seems reasonable to seek for an equation defining
${\bf {\bar x}}_{\cal F}$ which yields this solution for the spherically 
symmetric case. Furthermore, the wanted equation should be algebraic and linear
in ${\bf {\bar x}}_{\cal F}$ to ensure unique solutions, and it should involve 
${\bf {\bar a}}_{\cal F}$ since any deviation from spherical symmetry will be 
encoded into ${\bf {\bar a}}_{\cal F}$ and its spatial derivatives. By 
inspection of the spherically symmetric case, it turns out that it is possible 
to find an equation which has all the desired properties, namely
\eqa
{\Big [}{\bar a}_{{\cal F}{\mid}k}^k+c^{-2}{\bar a}_{{\cal F}k}
{\bar a}_{\cal F}^k{\Big ]}{\bar x}_{\cal F}^j-
{\Big [}{\bar a}_{{\cal F}{\mid}k}^j+c^{-2}{\bar a}_{{\cal F}k}
{\bar a}_{\cal F}^j{\Big ]}{\bar x}_{\cal F}^k-2{\bar a}_{\cal F}^j=0.
\ena
With ${\bf {\bar x}}_{\cal F}$ defined from equation (42) we are now able to 
define the 3-vector family ${\bf v}_t$. Expressed in a GTCS, the vector field 
family ${\bf v}_t$ by definition has the components [2]
\eqa
v^j_{(t)}{\equiv}{\bar y}_t{\bar x}_{\cal F}^j, {\qquad} 
v={\bar y}_t{\sqrt{{\bar h}_{(t)ks}{\bar x}_{\cal F}^k{\bar x}_{\cal F}^s}},
\ena
where $v$ is the norm of ${\bf v}_t$. Note that $v$ is required not to depend 
explicitly on $t$. So any dependence on $t$ of $v$ must be eliminated 
by setting $t=x^0/c$ where it occurs.

Now ${\bf g}_t$ is constructed algebraically from ${\bf {\bar g}}_t$ and $v$. 
To do that, as a first step we must compensate for the effects of the 
3-velocity $-{\bf v}_t$ of the ``static'' FOs in $({\cal N},{\bf {\bar g}}_t)$ 
on ${\bf {\bar g}}_t$. That is, the FOs in $({\cal N},{\bf {\bar g}}_t)$ are 
not ``at rest'', but move inwards with velocity $-{\bf v}_t$ compared to the 
FOs in $({\cal N},{\bf g}_t)$ when the local NKE is not taken into account. 
Said effects of $-{\bf v}_t$ are integrated into ${\bf {\bar g}}_t$ to begin 
with. To remove these effects, we include the effects of a compensating 
outwards 3-velocity ${\bf v}_t$ on coordinate line elements. In each tangent 
space, ${\bf v}_t$ yields a correction to spatial line intervals in the 
${\bf {\bar x}}_{\cal F}$-direction due to the radial Doppler effect, the 
correction factor being ${\Big (}{\frac{1+{\frac{v}{c}}}{1-{\frac{v}{c}}}}
{\Big )}^{1/2}$. There is also an inverse time dilation correction factor 
$(1-{\frac{v^2}{c^2}})^{1/2}$ to coordinate time intervals. There are no 
correction factors for spatial intervals normal to the 
${\bf {\bar x}}_{\cal F}$-direction. Second, we must transform the effects of 
the local NKE in $({\cal N},{\bf {\bar g}}_t)$ into correction factors to the 
coordinate line elements. In $({\cal N},{\bf {\bar g}}_t)$, the local NKE is 
described as an outwards 3-velocity ${\bf v}_t$ in the 
${\bf {\bar x}}_{\cal F}$-direction for each tangent space, and the effects of 
${\bf v}_t$, as seen by the FOs in $({\cal N},{\bf g}_t)$, are taken care of 
by using a pair of correction factors identical to those in step one.

To define transformation formulae, it is convenient to define the unit vector 
field ${\bf {\bar e}}_{\cal F}{\equiv}{\frac{t_0}{t}}{\bar e}_{\cal F}^s
{\frac{\partial}{{\partial}x^s}}$ and the corresponding covector field
${\bf {\bar {\omega}}}_{\cal F}{\equiv}{\frac{t}{t_0}}
{\bar {\omega}}_{{\cal F}s}dx^s$ along ${\bf {\bar x}}_{\cal F}$. Since it is in 
general not possible or practical to construct a GTCS where 
${\bf {\bar e}}_{\cal F}$ is parallel to one of the coordinate vector fields, 
one expects that the transformation formulae defining the transformation 
${\bf {\bar g}}_t{\rightarrow}{\bf g}_t$ should involve components of 
${\bf {\bar e}}_{\cal F}$ and ${\bf {\bar {\omega}}}_{\cal F}$. Requiring 
correspondence with the spherically symmetric case, one finds that [2]
\eqa
g_{(t)00}={\Big (}1-{\frac{v^2}{c^2}}{\Big )}^2{\bar g}_{(t)00},
\ena
\eqa
g_{(t)0j}={\Big (}1-{\frac{v^2}{c^2}}{\Big )}{\Big [}{\bar g}_{(t)0j}+
{\frac{t}{t_0}}{\frac{2{\frac{v}{c}}}{1-{\frac{v}{c}}}}
({\bar e}^s_{\cal F}{\bar N}_{(t)s}){\bar {\omega}}_{{\cal F}j}{\Big ]},
\ena
\eqa
g_{(t)ij}={\bar g}_{(t)ij}+{\frac{t^2}{t_0^2}}{\frac{4{\frac{v}{c}}}
{(1-{\frac{v}{c}})^2}}{\bar {\omega}}_{{\cal F}i}{\bar {\omega}}_{{\cal F}j}.
\ena
Note that we have eliminated any possible $t$-dependence of ${\bar N}_t$ in 
equations (44)-(46) by setting $t=x^0/c$ where it occurs. This implies that 
$N$ does not depend explicitly on $t$.

Some tensor fields are required to preserve their norm under the 
transformation defined in equations (44)-(46). One particular example of this 
is the transformation ${\bf {\bar F}}_t{\rightarrow}{\bf F}_t$ of the passive 
electromagnetic field tensor family ${\bf {\bar F}}_t$ in 
$({\cal N},{\bf {\bar g}}_t)$ to its counterpart ${\bf F}_t$ in 
$({\cal N},{\bf g}_t)$, where the latter enters into the Lorentz force 
law [4]. This suggests that the transformation 
${\bf {\bar g}}_t{\rightarrow}{\bf g}_t$ defined in equations (44)-(46) is 
only a particular case of a more general transformation. That is, 
transformations similar to ${\bf {\bar g}}_t{\rightarrow}{\bf g}_t$ should 
apply to any tensor field which norm is required to be unchanged when
${\bf {\bar g}}_t{\rightarrow}{\bf g}_t$. As an example we list the formulae
defining the transformation ${\bf {\bar Z}}_t{\rightarrow}{\bf Z}_t$ where
${\bf {\bar Z}}_t$ is a rank one tensor field family. (${\bf {\bar Z}}_t$ may,
e.g., be identified with a general 4-velocity vector family 
${\bf {\bar u}}_t$.) These formulae read
\eqa
Z_{(t)0}={\Big (}1-{\frac{v^2}{c^2}}{\Big )}{\bar Z}_{(t)0}, \qquad
Z_{(t)}^0={\Big (}1-{\frac{v^2}{c^2}}{\Big )}^{-1}{\bar Z}_{(t)}^0,
\ena
\eqa
Z_{(t)j}={\bar Z}_{(t)j}+{\frac{2{\frac{v}{c}}}{1-{\frac{v}{c}}}}
({\bar e}^s_{\cal F}{\bar Z}_{(t)s}){\bar {\omega}}_{{\cal F}j}, \qquad
{\hat Z}_{(t)}^j={\hat {\bar Z}}_{(t)}^j-
{\frac{2{\frac{v}{c}}}{1+{\frac{v}{c}}}}
({\bar {\omega}}_{{\cal F}s}{\hat {\bar Z}}_{(t)}^s){\bar e}_{\cal F}^j.
\ena
It is possible to find similar transformation formulae valid for higher rank 
tensor field families. For illustrative purposes we also list formulae valid 
for the transformation ${\bf {\bar W}}_t{\rightarrow}{\bf W}_t$, where
${\bf {\bar W}}_t$ is a rank two tensor field family. These formulae read
\eqa
W_{(t)00}={\Big (}1-{\frac{v^2}{c^2}}{\Big )}^2{\bar W}_{(t)00}, \qquad
W_{(t)}^{00}={\Big (}1-{\frac{v^2}{c^2}}{\Big )}^{-2}{\bar W}_{(t)}^{00},
\ena
\eqa
W_{(t)0j}&=&{\Big (}1-{\frac{v^2}{c^2}}{\Big )}{\Big [}{\bar W}_{(t)0j}+
{\frac{2{\frac{v}{c}}}{1-{\frac{v}{c}}}}({\bar e}^s_{\cal F}{\bar W}_{(t)0s})
{\bar {\omega}}_{{\cal F}j}{\Big ]}, \nonumber \\
{\hat W}_{(t)}^{0j}&=&{\Big (}1-{\frac{v^2}{c^2}}{\Big )}^{-1}
{\Big [}{\hat {\bar W}}_{(t)}^{0j}-{\frac{2{\frac{v}{c}}}{1+{\frac{v}{c}}}}
({\bar {\omega}}_{{\cal F}s}{\hat {\bar W}}_{(t)}^{0s})
{\bar e}_{\cal F}^j{\Big ]},
\ena
\eqa
W_{(t)ij}&=&{\bar W}_{(t)ij}+{\frac{2{\frac{v}{c}}}{(1-{\frac{v}{c}})^2}}
{\bar e}^k_{\cal F}({\bar {\omega}}_{{\cal F}i}
{\bar W}_{(t)kj}+{\bar W}_{(t)ik}{\bar {\omega}}_{{\cal F}j}), \nonumber \\
{\hat W}_{(t)}^{ij}&=&{\hat {\bar W}}_{(t)}^{ij}-{\frac{2{\frac{v}{c}}}
{(1+{\frac{v}{c}})^2}}{\bar {\omega}}_{{\cal F}k}({\bar e}_{\cal F}^i
{\hat {\bar W}}_{(t)}^{kj}+{\hat {\bar W}}_{(t)}^{ik}{\bar e}_{\cal F}^j).
\ena
These transformation formulae may easily be generalized to tensor field 
families of higher rank. Note that it is also possible to find formulae for 
the inverse transformations ${\bf Z}_t{\rightarrow}{\bf {\bar Z}}_t$,
${\bf W}_t{\rightarrow}{\bf {\bar W}}_t$ and similarly for tensor field
families of higher rank.

Since both ${\bf {\bar u}}_t$ and ${\bf {\bar F}}_t$ transform according to 
the rules (47)-(51), so must any 4-acceleration ${\bf {\bar a}}_t$ resulting
from the Lorentz force acting on charged matter. This implies that the norm of 
${\bf {\bar a}}_t$ must be invariant under the transformation. Since this 
result should apply to all 4-accelerations determined from non-gravitational 
forces and since such forces may in particular vanish, we may deduce that 
{\em geodesic motion in $({\cal N},{\bf {\bar g}}_t)$ implies geodesic motion 
in $({\cal N},{\bf g}_t)$.} That is, any inertial observer in
$({\cal N},{\bf {\bar g}}_t)$ must be inertial in $({\cal N},{\bf g}_t)$ as
well. Note that the 4-acceleration field ${\bf {\bar a}}_{\cal F}$ does not
in general transform according to the rules (47)-(48) since 
${\bf {\bar a}}_{\cal F}$ is determined from the requirement that the FOs must
move normal to the FHSs rather than from some non-gravitational force acting on
the FOs.

The vector field family ${\bf v}_t$ is an extra dynamical field constructed 
from ${\bf {\bar g}}_t$ and the distinguished foliation of 
$({\cal N},{\bf {\bar g}}_t)$ into spatial hypersurfaces. This means that
${\bf v}_t$ is not an {\em independent} dynamical field; however the dynamics
represented by ${\bf v}_t$ is {\em implicit} inasmuch as it is not explicitly 
coupled to matter fields. Moreover, since it can be calculated on each FHS
from already known quantities there, ${\bf v}_t$ does not represent any
{\em propagating} dynamical degree of freedom. That is, unlike 
${\bf {\bar h}}_t$, which dynamical evolution is present explicitly in the 
field equations via the extrinsic curvature ${\bf {\bar K}}_t$, no such 
explicit presence exists for ${\bf v}_t$.

We close this section by emphasizing that ${\bf g}_t$ is the ``physical'' 
metric family in the sense that ${\bf g}_t$ should be used consistently when 
comparing predictions of QMR to experiments (and in particular to observations 
involving the equations of motion). That is, any laws given in terms of 
${\bf {\bar g}}_t$ and its associated connection, e.g. the local conservation 
laws defined in equation (35), are not the ``physical'' laws; those must always
be in terms of ${\bf g}_t$ and its associated connection when comparing 
directly to experiment. Nevertheless it is sometimes necessary to use the laws 
in terms of ${\bf {\bar g}}_t$ and its associated connection. For example, to 
be able to calculate ${\bf {\bar g}}_t$ it is in general necessary to use the 
local conservation laws (35). But as long as one is aware of the correct 
relationship between laws and observables this should not represent any 
problem.
\subsection{Comparing theory to experiment}
To determine whether or not quasi-metric theory might be viable, we must try to 
confront its predictions with data from modern relativistic gravitational 
experiments. But the two first questions are if it has a consistent weak-field 
limit and how this limit relates to the Newtonian limit of metric theories. 
Now Newtonian theory should have a correspondence with the {\em metric} 
part of quasi-metric theory, and not with the non-metric part, which has no 
Newtonian counterpart. Thus, even if the local NKE can be neglected at the 
Newtonian level of precision, there is no reason to think that this also 
applies to the global NKE. Therefore, a more useful approximation than the 
traditional Newtonian limit can be made by taking the weak-field limit 
of equation (13), but such that the global spatial scale factor 
${\frac{t}{t_0}}$ is included. That is, in this ``quasi-Newtonian'' limit, 
the FHSs are taken as flat, but non-static since the global NKE is included. 
Besides, all the local NKE is ignored in the quasi-Newtonian limit so that 
${\bf {\bar g}}_t={\bf g}_t$ (and we can thus drop the bar labels if 
convenient). 

Taking into account the above discussion, we may now write down the weak-field
limit of equation (13) at the Newtonian level of precision. To do that, we 
estimate the smallness of the terms to be of the same order as that of the 
small quantity ${\frac{w}{c}}$ to some power, where $w$ is the typical speed of
the (gravitating) matter with respect to the FOs. For an isolated system
and weak gravitational fields, to Newtonian accuracy the FOs can be chosen to 
be at rest with respect to a suitable GTCS using Cartesian coordinates (see 
section 3.3). The quasi-Newtonian metric family then has the components
\eqa
{\bar g}_{(t)00}=-1+2c^{-2}U(x^{\mu})+O(4), \quad
{\bar g}_{(t)i0}={\bar g}_{(t)i0}=0+O(3), {\quad} 
{\bar h}_{(t)ij}={\frac{t^2}{t_0^2}}{\delta}_{ij}+O(2),
\ena
where $-U(x^{\mu})$ is the Newtonian potential. Note that equation (52) is 
consistent with the general metric family (13) since to Newtonian accuracy, 
we can neglect any contribution to ${\bar h}_{(t)ij}$ from $c^{-2}U(x^{\mu})$ 
as this term is of $O(2)$. The quasi-Newtonian form (52) of the metric 
family is useful since it takes sufficiently care of the effects of the global 
NKE for weak gravitational fields and slow motions. Moreover, the traditional 
Newtonian metric form can be recovered just by setting the factor 
${\frac{t}{t_0}}$ equal to unity in equation (52).

Since the weak-field approximation of the extrinsic curvature tensor family 
${\bf {\bar K}}_t$ is at least of $O(3)$ or higher, it may be neglected in the 
field equations and in the local conservation laws at the Newtonian level of 
precision. Then we see that equation (20) yields Newton's field equation 
whereas equations (23) and (34) become vacuous for sufficiently weak 
gravitational fields. Moreover, it is straightforward to show that for a 
perfect fluid source given by equation (39), to Newtonian accuracy each of 
equations (36)-(37) corresponds to the counterpart Euler equation valid for 
Newtonian fluid dynamics.

The next level of precision beyond the Newtonian limit is the post-Newtonian
approximation (applied to isolated gravitational systems). At this level of 
precision, the motion of the FOs cannot be ignored and nor can the local NKE. 
This means that any attempt to construct a general weak-field approximation 
formalism for quasi-metric gravity (along the lines of the parametrized 
post-Newtonian (PPN) formalism valid for metric theories of gravity) will meet 
some extra complications. This also implies that it is not a good idea to try 
to apply the standard PPN-formalism to our quasi-metric theory. There are 
several reasons for this; one obvious reason is that the PPN-formalism neglects
the non-metric aspects of QMR. This means that any PPN-analysis of our field 
equations is limited to their metric approximations. But even these metric 
approximations are not suitable for a standard PPN-analysis since the resulting
PPN-metric ${\bf {\bar g}}$ (the $t$-labels are omitted everywhere when dealing
with metric approximations) is not the one to which experiments are to be 
compared, and ${\bf {\bar g}}$ will {\em not} have an acceptable set of 
PPN-parameters according to metric theory. For example, for metrically static 
systems, a PPN-analysis of our field equations yields the PPN-parameters 
${\gamma}=-1$ and ${\beta}=0$; both values are totally unacceptable for any 
viable metric theory. 

As discussed in section 3.3, for a sufficiently small isolated system, the 
global curvature of space may be neglected to a good approximation. 
Furthermore, an approximately global cosmic frame with an associated 
approximately global time function may be chosen such that the barycentre of 
the system is taken to be at rest with respect to this frame (which may be 
identified with the standard PPN coordinate system). The PPN-approximations of
the field equations may then be transformed to this frame. But they will not be
invariant under this transformation since said frame represents an alternative 
foliation of space-time into spatial hypersurfaces. That is, in QMR there 
should be ``preferred frame''-effects somewhat resembling those covered by the 
PPN-formalism, and with the condition ${\gamma}=-1$. However, one must be 
careful not to interpret said effects as due to a variable gravitational 
``constant'' on top of Newtonian theory (as is done in a standard PPN-analysis 
[1, 6]); this would be inconsistent with quasi-metric gravity. As a result, the
detectability of any ``preferred frame''-effects should be significantly more 
subtle for quasi-metric gravity than for metric theories of gravity.

Moreover, finding a more complete set of PPN-parameters for ${\bf {\bar g}}$ 
turns out to be problematic since the relationships assumed to hold between 
said parameters in metric theories will not necessarily hold for quasi-metric 
gravity. This typically leads to inconsistencies. For example, the differences 
between QMR and metric gravity regarding the implementation of the EEP show up 
via the local conservation laws (35) since even in the metric approximation, 
these laws are different from their counterparts in standard metric theory. 
This means that any constraints on the PPN-parameters deduced from integral 
conservation laws [1] will not necessarily hold in QMR.

All of the above implies that any further analysis of the field equations to 
find a complete set of PPN-parameters for ${\bf {\bar g}}$ would be rather
pointless. Furthermore, when one attempts to construct a ``physical'' 
PPN-metric ${\bf g}$ from ${\bf {\bar g}}$ in the manner discussed in the 
previous section, one gets more complications. In particular, the 
transformation ${\bf {\bar g}}{\rightarrow}{\bf g}$ would turn the
PPN-parameters for ${\bf {\bar g}}$ into scalar fields rather than new 
constants, reducing the usefulness of a potential PPN-metric ${\bf g}$. 
Besides, the isotropic PPN coordinate system is not invariant 
under the transformation ${\bf {\bar g}}{\rightarrow}{\bf g}$. That is, 
isotropic coordinates for the metric ${\bf g}$ are different from the isotropic 
coordinates one started out with in the first place when solving the field 
equations! The reason for this is, of course, that the construction of 
isotropic coordinates depends on the metric. But the main problem here is that 
the PPN-formalism does not tackle properly the construction of ${\bf g}$ from 
${\bf {\bar g}}$. That is, the PPN-formalism exclusively handles explicit 
gravitational dynamics represented by field equations and neglects the possible
existence of implicit dynamics as represented by the velocity field ${\bf v}$. 
Thus the bottom line is that a standard PPN-analysis, even limited to metric 
approximations of QMR, will fail.

So fact is that, to be able to compare the predictions of our theory to
gravitational experiments performed in the solar system in a satisfactory way,
a separate weak-field expansion similar to the PPN-formalism should be 
developed. And since such a formalism is lacking at this point in time, it 
is not yet clear whether or not quasi-metric theory is viable. However, 
if a separate formalism is developed it should have some correspondence with
the PPN-formalism to answer this question (minimizing the need for reanalyzing
weak-field experiments within the new framework). But we may still calculate 
specific solutions with high symmetry to get an idea how the quasi-metric 
theory compares to GR. In particular, in the metric approximation we may 
calculate the exact counterpart to the Schwarzschild case of GR. That is, in 
Schwarzschild coordinates, for the static, spherically symmetric vacuum 
exterior to an isolated source, the field equations in the metric 
approximation yield the (unique) solution [2] 
\eqa
{\overline{ds}}^2={\Big (}{\sqrt{1+({\frac{r_{\rm s}}{2r}})^2}}-
{\frac{r_{\rm s}}{2r}}{\Big )}^2{\Big (}-(dx^0)^2+
[1+({\frac{r_{\rm s}}{2r}})^2]^{-1}dr^2{\Big )}+r^2d{\Omega}^2.
\ena
Here, $d{\Omega}^2{\equiv}d{\theta}^2+{\sin}^2{\theta}d{\phi}^2$ and $r_{\rm s}$ 
is the generalized Schwarzschild radius defined by
\eqa
r_{\rm s}{\equiv}{\frac{2M^{\rm (EM)}G^{\rm B}}{c^2}}+
{\frac{2M^{\rm (MA)}G^{\rm S}}{c^2}}, \qquad
M^{\rm (EM)}{\equiv}c^{-2}{\int}{\int}{\int}{\bar N}
(T^{\rm (EM)}_{{\bar {\perp}}{\bar {\perp}}}+
{\hat T}^{{\rm (EM)}i}_{{\ }i})d{\bar V}, \nonumber \\
M^{\rm (MA)}{\equiv}c^{-2}{\int}{\int}{\int}{\bar N}
(T^{\rm (MA)}_{{\bar {\perp}}{\bar {\perp}}}+
{\hat T}^{{\rm (MA)}i}_{{\ }i})d{\bar V}.
\ena
In equation (54), $M^{\rm (EM)}+M^{\rm (MA)}$ is the total dynamical (Komar) mass 
of the source and the integrations are taken over the FHS. (The particular 
form of $M^{\rm (EM)}$ and $M^{\rm (MA)}$ follows directly from the field equations 
applied to the interior of a spherically symmetric, static source when 
extrapolated to the exterior solution given by equation (53). See reference
[5] for details.) We notice that the metric (53) has no event
horizon, but that there is a curvature singularity at the origin. 

Furthermore, since ${\bf {\bar x}}_{\cal F}=r{\frac{\partial}{{\partial}r}}$ for
the spherically symmetric case, we easily find from equations (18) and (43) 
that
\eqa
v={\frac{r_{\rm s}c}{2r{\sqrt{1+({\frac{r_{\rm s}}{2r}})^2}}}}.
\ena
Then, using equations (44) and (46), we get
\eqa
ds^2&=&-[1+({\frac{r_{\rm s}}{2r}})^2]^{-2}{\Big (}{\frac{r_{\rm s}}{2r}}+
{\sqrt{1+({\frac{r_{\rm s}}{2r}})^2}}{\Big )}^{-2}(dx^0)^2 \nonumber \\
&{ }&+[1+({\frac{r_{\rm s}}{2r}})^2]^{-1}{\Big (}{\frac{r_{\rm s}}{2r}}+
{\sqrt{1+({\frac{r_{\rm s}}{2r}})^2}}{\Big )}^{2}dr^2+r^2d{\Omega}^2 
\nonumber \\ &{ }&=-(1-{\frac{r_{\rm s}}{r}}+{\frac{3}{8}}
{\frac{r_{\rm s}^3}{r^3}}+{\cdots})(dx^0)^2+(1+{\frac{r_{\rm s}}{r}}+
{\frac{r_{\rm s}^2}{4r^2}}+{\cdots})dr^2+r^2d{\Omega}^2.
\ena
We see that this metric has no event horizon either, and that it is consistent 
with the four ``classical'' solar system tests. Notice that this consistency is 
due to the existence of the implicit gravitational dynamics represented by the 
vector field ${\bf v}$.

It is important to notice that the metric (56) is only a metric approximation 
yielding correspondences between QMR and GR. But we may go beyond the
metric approximation and include the effects of the non-metric part of QMR 
in the spherically symmetric case. This is done in references [2] and [5] 
where it is shown that the quasi-metric theory predicts that the size of the 
solar system increases according to the Hubble law, but in a way such that the 
trajectories of non-relativistic test particles are not unduly affected. 
However, this prediction has a number of observable consequences which are 
seen and in good agreement with predictions calculated from QMR [5, 7]. In 
particular, the prediction that the solar system expands according to the 
Hubble law provides a natural explanation [7] of the apparently ``anomalous'' 
acceleration of some distant spacecraft as inferred from radiometric data [8].

We conclude that, even if the PPN-formalism does not apply to QMR and that this
makes the predictions of QMR more difficult to test against experiment, some 
of the non-metric aspects of QMR seem to agree well with observations. This
represents a challenge for GR and other metric theories just as much as the
successes of GR represent a challenge for any alternative theory of gravity.
But it is a mathematical fact that metric theories are unable to handle the
non-metric aspects of QMR in a geometrical manner, making it impossible
to calculate any of these effects from first principles in metric gravity.
\section{Quasi-metric cosmology}
\subsection{General predictions}
Cosmology as done in QMR is radically different from any possible approach to 
the subject based on some metric theory of gravity. The main reason for this 
is, of course, that in QMR the expansion of the Universe is not interpreted 
as a ``kinematical'' phenomenon (in a general sense of the word). Rather, by 
construction the cosmic expansion is a prior-geometric property of quasi-metric
space-time itself. This means that any concept of the Universe as a purely 
gravitational dynamical system simply is not valid in QMR. Consequently, many 
of the problems encountered in traditional cosmology do not exist in 
quasi-metric cosmology. For example, in QMR the expansion history of the 
Universe does not depend on its matter density, so there is no flatness 
problem. Due to the coasting expansion no horizon problem exists either, nor is
there any need for a cosmological constant. Thus QMR yields some cosmological 
predictions from first principles, without the extra flexibility represented by
the existence of a set of cosmological parameters. In QMR there may be 
cosmological problems not encountered in metric gravity, however. 

The lack of any sort of cosmic dynamics in QMR is realized mathematically by 
the fact that no quasi-metric counterparts to the Friedmann-Robertson-Walker 
(FRW) models exist [2]. However, one possible cosmological model with 
isotropic FHSs is the toy model given by the metric family
\eqa
ds_t^2={\overline{ds}}_t^2&=&-(dx^0)^2+({\frac{t}{t_0}})^2{\Big (}
{\frac{dr^2}{1-{\frac{r^2}{(ct_0)^2}}}}+r^2d{\Omega}^2{\Big)} \nonumber \\
&=&-(dx^0)^2+(ct)^2{\Big (}d{\chi}^2+{\sin}^2{\chi}d{\Omega}^2{\Big )},
\ena
which represents an empty universe. This is a family of 
${\bf S}^3{\times}{\bf R}$ space-time metrics, and it is easy to check that it
satisfies the field equations without sources and also equations (17), (18). 
Besides this empty model, it is possible to have a toy cosmological model where
the Universe is filled with an isotropic null fluid. In this case one finds 
solutions conformal to the solution (57), with a conformal factor equal to
${\bar N}_t^2$. Moreover, up to a constant factor, we find that 
${\bar N}_t={\exp}[-K{\frac{(x^0)^2}{(ct)^2}}]$ (where $K$ is a constant 
depending on the fluid density) from the field equations. Such solutions also 
satisfy equations (36), (37). But since ${\bar N}_t$ is constant on the FHSs in
these models, we may transform the resulting ${\bf g}_t$ into the metric family 
shown in equation (57) by setting ${\bar N}_t={\exp}[-K]$ and doing trivial 
re-scalings of the time coordinate ${\bar N}_tx^{0}{\rightarrow}x^{0}$, the 
radial coordinate ${\bar N}_tr{\rightarrow}r$, and of the global time function
${\bar N}_tt{\rightarrow}t$. We may thus use the solution (57) in the equations
of motion even for the case when the Universe is filled with a null fluid.
It is also possible to find isotropic null fluid models where there is local 
creation of null particles. In such models ${\bar N}_t$ will effectively 
depend on $t$ in $({\cal N},{\bf {\bar g}}_t)$, and equation (38) will be
violated.

Now one peculiar aspect of QMR is that gravitationally bound bodies made of
ideal gas and their associated gravitational fields are predicted to expand 
according to the Hubble law [2, 5]. That is, if net energy transfer between 
electromagnetic fields and material particles in addition to fluid-dynamical
instabilities due to the expansion can be neglected (there are no such 
instabilities for an ideal gas), measured in atomic units, linear sizes within 
a gravitationally bound system will increase as the scale factor, i.e., 
proportionally to $t$. Notice that this is valid even for the 
quantity $c^{-2}G^{\rm B}M^{\rm (EM)}_t+c^{-2}G^{\rm S}M^{\rm (MA)}_t$ (where 
$M_t^{\rm (EM)}$ and $M_t^{\rm (MA)}$ are active masses), which has the dimension of
length. On the other hand it is a prediction of QMR that, except for a global 
cosmic attenuation not noticeable locally, the passive electromagnetic field 
is unaffected by the global cosmic expansion [4]. This means that there is no
reason to expect that atoms or other purely quantum-mechanical systems should 
participate in the cosmic expansion. (To clarify the effect of the expansion 
on atoms it would be necessary to perform calculations involving quantum 
fields in quasi-metric space-time.)

A universe filled with an isotropic fluid consisting of material particles is
not possible in QMR. That is, for any toy model universe filled with a perfect
fluid source described by an equation of state different from 
${\varrho}_{\rm m}c^2=3p$, ${\bar N}_t$ must necessarily vary in space as well 
as in time. Besides, it is not expected that these cases should yield exact
solutions conformal to the solution (57). However, the more relativistic the 
fluid, the closer its equation of state will be to that of a null fluid. This 
means that one expects the deviations from isotropy to be very small in the 
early Universe, so that the isotropic null fluid solution should be a good 
approximation. On the other hand, one expects increasing deviations from 
isotropy when the primaeval cosmic matter cools and eventually becomes 
non-relativistic. Then said null fluid solution ceases to be a good approximate
solution, and one must in principle find the solutions corresponding to the
cosmic equation of state. These solutions cannot be isotropic. That is, one 
expects that gravitationally induced deviations from isotropy must necessarily 
increase with cosmic epoch in QMR, as a direct consequence of the transition of
the cosmic matter from highly relativistic to non-relativistic. Thus no 
fine-tuning will be necessary to get a clumpy universe from a near-isotropic 
beginning.

A valid interpretation of equation (57) is that fixed operationally defined
atomic units vary with epoch $t$ in such a way that atomic length units shrink
when $t$ increases. This means that no matter can have been existing from the 
beginning of time since atomic length units increase without bound in the 
limit $t{\rightarrow}0$. Consequently, we may take an empty model described by 
equation (57) as an accurate cosmological model in this limit. Thus QMR yields
a natural description of the beginning of time (with no physical singularity)
where all big bang models fail (since big bang models are not valid for 
$t=0$). But an empty beginning of the Universe means that one needs a working 
matter creation mechanism. Thus it is natural to suggest something analogous 
to particle creation by the expansion of the Universe in traditional big bang 
models. That is, in the very early Universe the global NKE is so strong that 
non-gravitational quantum fields cannot be treated as localized to sufficient 
accuracy, so one should get spontaneous pair production from excitations of 
vacuum fluctuations of such quantum fields (violating equation (38)). 
Moreover, newly created material particles should induce tiny gravitational 
perturbations which will grow when the Universe cools. The details of these
suggestions have not been worked out. However, any hope that QMR may represent 
a complete framework for relativistic physics depends on if the mathematical
details of a matter creation mechanism can be developed.

Even if models of the type (57) are not accurate for the present epoch, we may 
still use it to illustrate some of the properties of a cosmological model 
where the expansion is non-kinematical. That is, the linear dependence of the 
scale factor on $ct$ and the global positive curvature of space are valid 
predictions of any quasi-metric cosmological model, so even if a more realistic
model with non-isotropic matter density does represent a deviation from 
equation (57), we may still use equation (57) in combination with the 
equations of motion to deduce some general features of quasi-metric cosmology. 
In particular, it is easy to derive the usual expansion redshift of momentum 
for decoupled massless particle species from equation (57). To do that, use 
the coordinate expression for a null path in the ${\chi}$-direction as 
calculated from equation (57) and the equations of motion. The result is [2]
\eqa
{\chi}(t)={\chi}(t_0)+{\ln}{\frac{t}{t_0}},
\ena
and a standard calculation using equation (58) yields the usual expansion 
redshift formula. Also the corresponding time dilation follows from equation 
(58).

In standard cosmology, the cosmic expansion affects the momenta of a photon and
of an extremely relativistic material particle the same way, meaning that the 
temperature evolution of an extremely relativistic plasma in thermal 
equilibrium goes as the inverse of the scale factor (neglecting potential
heating effects coming from net particle-antiparticle annihilation). On the 
other hand, in QMR the speed $w$ of any inertial material point particle with 
respect to the FOs is unaffected by the global NKE [2]. This means that the 
plasma temperature and scale factor evolutions will be differently related from
their counterparts in metric theory as long as the number of material particles
in thermal equilibrium with the cosmic photon plasma is not negligible. The 
effect of this on primordial nucleosynthesis is not known. However, it seems
that coasting universe models in metric gravity are somewhat consistent with 
primordial nucleosynthesis of $^4$He [9], but apparently there is no natural 
way to match the observed abundances of deuterium and $^3$He. Due to the 
similarity of the time scales involved, it is not likely that QMR can do
significantly better than said coasting models regarding primordial 
nucleosynthesis.

Over the last few years, a ``concordance'' big bang model has emerged from the 
observational determination of standard cosmological parameters. Key
observational constraints on these cosmological parameters mainly come from two
different types of data; i.e., from supernovae at cosmological distances (see 
the next section) and from analysis of temperature fluctuations in the cosmic 
microwave background (CMB). The concordance model predicts that the Universe 
should be nearly spatially flat and filled with exotic matter (``dark 
energy'', possibly in the form of a cosmological constant) dominating the 
dynamics of the Universe, causing the cosmic expansion to accelerate. Since a 
universe dominated by dark energy raises some rather deep (and potentially 
unanswerable) questions concerning its nature, it could be argued that the 
inferred existence of the preposterous dark energy might be an artefact due to
analyzing observational data within an incorrect space-time framework. To 
explore this possibility, the CMB data should be reanalyzed within the 
quasi-metric framework. In particular this should be done for the data 
obtained from the Wilkinson Microwave Anisotropy Probe (WMAP). These data 
show at least one unexpected feature; namely that the temperature angular 
correlation function lacks power on angular scales greater than about 
60$^{\circ}$ [10]. This may possibly have a natural explanation within the 
quasi-metric framework since any quasi-metric universe is closed and 
``small''; i.e., its spatial curvature scale should represent a natural 
cut-off for fluctuations. Moreover, since no counterpart to the Friedmann 
equation exists in QMR, there are no dynamical restrictions on the data ruling
out a universe with scale factor close to the size of the observable Universe.
This may be compared to the metric framework, where a closed universe cannot 
be too small and have a trivial topology since this will be inconsistent with 
the value of the Hubble parameter obtained from other, independent 
observations (see, e.g., reference [11] for a further discussion of this 
point).
\subsection{QMR and type Ia supernovae}
Newtonian stars for which the equation of state takes the form
$p{\propto}{\varrho}_{\rm m}^{\gamma}$, ${\gamma}>{\frac{6}{5}}$, are called 
{\em Newtonian polytropes} [12]. In quasi-metric theory it is possible to 
model Newtonian polytropes by taking Newtonian limits of the relevant equations
but such that the $t$-dependence remains. That is, according to quasi-metric 
theory, the cosmic expansion is predicted to induce fluid-dynamical 
instabilities in polytropes consisting of degenerate matter [2, 5]. But if the 
dynamical effects on the gravitational field coming from instabilities 
can be neglected, one may solve the field equations for each epoch $t$ 
assuming that the polytrope is in hydrostatic equilibrium. It is then possible
to show [2] that the usual analysis of Newtonian polytropes [12] 
applies, but with with a variable $G^{\rm S}$, i.e. 
$G^{\rm S}{\rightarrow}G^{\rm S}_t{\equiv}{\frac{t}{t_0}}G^{\rm S}$
(any dependence on a variable $G^{\rm B}$ is assumed to be negligible).

Of particular interest are Newtonian polytropes for which ${\gamma}=
{\frac{4}{3}}$, since such stars are models for Chandrasekhar mass white 
dwarfs (WDs). From reference [12] we easily find that the (passive) mass 
$m_{\rm c}$ and physical radius ${\cal R}_{\rm c}$ of such WDs (with identical
central mass densities) depend on epoch such that, respectively, 
$m_{\rm c}(t)=({\frac{t_0}{t}})^{3/2}m_{\rm c}(t_0)$ and 
${\cal R}_{\rm c}(t)=({\frac{t_0}{t}})^{1/2}{\cal R}_{\rm c}(t_0)$. Since
Chandrasekhar mass WDs are believed to be progenitors of type Ia supernovae,
one may expect that any cosmic evolution of Chandrasekhar mass WDs should 
imply a systematic luminosity evolution of type Ia supernovae over cosmic time 
scales. However, such a luminosity evolution would be inconsistent with their 
usage as standard candles when determining the cosmological parameters in 
standard cosmology: luminosity evolution could have serious consequences for 
an interpretation of the supernova data in terms of an accelerating cosmic 
expansion indicating a non-zero cosmological constant [13, 14]. 

Now quasi-metric theory predicts that the cosmic expansion does neither 
accelerate nor decelerate. Moreover, according to quasi-metric theory, the
Chandrasekhar mass decreases with epoch and this means that type Ia supernovae 
may be generated from cosmologically induced collapse of progenitor WDs. But
the consequences for type Ia supernova peak luminosities due to the predicted 
evolution of progenitor WDs are not clear. Since the luminosity of type Ia
supernovae comes from ${\gamma}$-disintegration of unstable nuclear species
(mainly $^{56}$Ni) synthesized in the explosion, this luminosity could depend
critically on the conditions of the nuclear burn. That is, the detailed nuclear
composition synthesized in the explosion might depend on the local acceleration
of gravity experienced by the burning front (which should depend on the value 
of $G^{\rm S}$). Other critical factors might be supernova progenitor mass and 
chemical composition [15] (see below). Also the presence of more massive ejecta 
during the explosion could have an influence on supernova luminosities and 
light curves. 

The effects of a varying gravitational ``constant'' on type Ia supernova
luminosities have been studied elsewhere [16, 17]. In these papers, it is 
assumed that the peak luminosity $L$ is proportional to the synthesized 
mass of $^{56}$Ni which again is assumed to be proportional to the 
Chandrasekhar mass. Moreover, using a toy model of the supernova explosion, 
the dependences of $L$, and the intrinsic time scale ${\tau}_{\rm s}$ of the 
explosion, on $G^{\rm S}$ have been deduced (neglecting radioactive heating). 
The results are
\eqa
L(t){\propto}{\Big (}{\frac{G^{\rm S}(t)}{G^{\rm S}(t_0)}}{\Big )}
^{-{\frac{3}{2}}}, \qquad
{\tau}_{\rm s}(t){\propto}{\Big (}
{\frac{G^{\rm S}(t)}{G^{\rm S}(t_0)}}{\Big )}^{-{\frac{3}{4}}}.
\ena
We see that if both of these luminosity and intrinsic time scale evolutions 
were valid within the quasi-metric framework, we would have been forced to 
deduce that $L(z){\propto}(1+z)^{\frac{3}{2}}$ and ${\tau}_{\rm s}(z){\propto}
(1+z)^{\frac{3}{4}}$. This means that, rather than intrinsically fainter, 
ancient supernovae would have been predicted to be intrinsically 
{\em brighter} than today's in addition to displaying broader light curves, 
contrary to observations. But it is by no means obvious that the toy model 
yielding the above luminosity and intrinsic time scale evolutions is
sufficiently realistic. Rather, one should include all aspects of progenitor 
evolution on type Ia supernovae before deducing such evolutions. In particular,
the amount of $^{56}$Ni synthesized in the explosion and thus the supernova 
luminosity will depend on the C/O ratio present in the progenitor WD. In 
general, more massive progenitor WDs will have smaller C/O ratios than less 
massive ones due to different conditions in the He-burning stages of their
progenitor stars [18]. A smaller C/O ratio implies that a smaller percentage 
of $^{56}$Ni should be produced relative to other explosion products [18]. 
Besides, a smaller C/O ratio will affect the energetics of the explosion such 
that ejecta velocities and thus supernova size will be smaller at any given 
time after the explosion [18]. Since these effects may be important for 
supernova luminosity evolution, it is possible that ancient supernovae in fact 
could be intrinsically dimmer than today's, despite their being more massive. 
But any investigation of this possibility requires detailed numerical 
simulations, so before such have been performed it is not possible to say 
whether or not the predictions from quasi-metric theory are consistent with 
the data.
 
However, what we {\em can} easily do is to see if it is possible to construct 
a simple luminosity evolution which, in combination with the cosmological toy 
model (57), yields a reasonable fit to the supernova data. That is, we may try
a luminosity evolution of the source of the form 
\eqa
L_{\rm qmr}=L_{\rm std}({\frac{t_0}{t}})^{\epsilon}=
L_{\rm std}(1+z)^{-{\epsilon}}, 
\ena
where $L_{\rm std}$ is a fixed standard luminosity, and see if the data are 
well fitted for some value(s) of ${\epsilon}$. To check this, we use the 
postulated luminosity evolution to plot apparent magnitude $m_{\rm qmr}$ 
versus redshift for type Ia supernovae. The easiest way to compare this to 
data, is to calculate the predicted difference between the quasi-metric model 
(with source luminosity evolution) and a FRW model where the scale factor 
increases linearly with epoch, namely the ``expanding Minkowski universe'' 
given by a piece of Minkowski space-time:
\eqa
ds^2=-(dx^0)^2+(x^0)^2{\Big (}d{\chi}^2+{\sinh}^2{\chi}d{\Omega}^2{\Big )}.
\ena
The difference in apparent magnitude ${\Delta}m$ between the two models (as a 
function of redshift $z$) can be found by a standard calculation. The result is
\eqa
{\Delta}m{\equiv}m_{\rm qmr}-m_{\rm min}&=&
2.5{\log}_{10}{\Big [}{\sin}^2{\{}{\ln}(1+z){\}}{\Big ]} \nonumber \\
&&-5{\log}_{10}{\Big [}{\sinh}{\{}{\ln}(1+z){\}}{\Big ]}-
2.5{\log}_{10}{\frac{L_{\rm qmr}}{L_{\rm min}}},
\ena
where $L_{\rm qmr}/L_{\rm min}$ represents the luminosity evolution of the 
source in our quasi-metric model relative to no luminosity evolution in an
empty FRW model. One may then find the relation $m_{\rm qmr}(z)$ from
equation (62) and the relation $m_{\rm min}(z)$ graphically shown in 
reference [13], and then compare to data. One finds that the quasi-metric 
model is quite consistent with the data for values of ${\epsilon}$ of about 
$0.5$. That is, for ${\epsilon}$ near $0.5$, ${\Delta}m$ has a maximum at 
$z{\approx}0.5$; for higher redshifts ${\Delta}m$ decreases (and eventually 
becomes negative for $z$ larger than about $1.2$). In standard cosmology this 
behaviour would be interpreted as evidence for an era of cosmic deceleration 
at high $z$.

We conclude that quasi-metric cosmology combined with a simple luminosity
evolution seems to be consistent with the data but that a much more 
detailed model should be constructed to see if the postulated luminosity
evolution has some basis in the physics of type Ia supernovae.

From the above, we see that the assertion that type Ia supernovae can be used 
as standard cosmic candles, independent of cosmic evolution, is a 
model-dependent assumption. But the fact is that models for which this holds 
fail to explain the effects of the cosmic expansion seen in the solar 
system [5, 7]. Thus, any interpretation of the supernova data 
indicating that the expansion of the Universe is accelerating should be met 
with some extra scepticism.
\section{Conclusions}
In many ways, any theory of gravity compatible with the quasi-metric framework
must be fundamentally different from metric theories of gravity since 
quasi-metric space-time is not modeled as a pseudo-Riemannian manifold. The 
most obvious of these differences is the existence of a non-metric sector and 
the fact that it directly influences the equations of motion. This means that 
the existence of a non-metric sector may be inferred from data on test particle 
motion. In fact, non-metric effects on test particle motion in weak 
gravitational fields can be tested against experiment rather independently of 
any systematic weak-field expansion for the metric sector. And the status so 
far is that it seems like non-metric effects are seen in good agreement with 
predictions [5, 7].

The geometrical structure of quasi-metric space-time also gives QMR some
conceptual advantages over metric theory. For example, the simple causal 
structure of QMR leaves no room for event horizons. Thus potential vexing 
questions concerning the nature of black holes and space-time singularities do
not exist in QMR. Besides, quasi-metric space-time is constructed to yield
maximal predictive power. This shows up most clearly in cosmology, since both 
the predicted global shape and curvature of the Universe in addition to its 
expansion history are basic features of QMR and not adjustable. This is in 
contrast to metric theory, where essentially arbitrary cosmic initial 
conditions and a number of cosmological parameters are available, making it 
much more flexible than QMR. Should QMR turn out to survive confrontation with 
cosmological observations, the vulnerability of quasi-metric predictions would 
give strong support to QMR. 

But quasi-metric gravity, even in its metric sector, is different from those
metric theories suitable for a standard PPN-analysis. One reason for this is 
that quasi-metric gravity contains a (not independent) {\em implicit} 
dynamical gravitational vector field family ${\bf v}_t$ not coupled explicitly 
to matter. Such a feature does not have any correspondence with the standard 
PPN-framework. Another reason is that relationships between PPN-parameters 
assumed to hold for metric gravity, will not necessarily hold for quasi-metric 
gravity. One expects that this will lead to inconsistencies. Unfortunately, the
lack of a weak-field expansion formalism (having some necessary correspondence 
with the PPN-framework) makes it harder to test the metric aspects of QMR. 
However, a weak-field expansion scheme is not needed to see that the geometric 
structure of quasi-metric space-time is consistent with no Nordtvedt effect. In
other words, the quasi-metric theory of gravity presented in this paper should 
fulfil the Gravitational Weak Equivalence Principle (GWEP). In fact, many 
experiments testing the validity of the SEP actually test the GWEP. Thus the 
fact that QMR violates the SEP needs not be fatal.

But there are other crucial observational tests which QMR has to survive. For 
example, the detailed spectrum of temperature fluctuations in the CMB must be 
consistent with predictions coming from QMR. Another task is to construct a 
model of primordial nucleosynthesis in QMR. Predictions coming from such a 
model must be at least be consistent with the observed abundance of $^4$He. 
For abundances of other light nuclear species, e.g., deuterium and $^3$He, 
there must be found some other natural explanations if the predicted 
primordial abundances do not match observations. So there is much further work 
to be done before we can know whether or not QMR is viable. However, 
observations do seem to confirm the existence of a non-metric sector. This 
suggests that metric theory is wrong so QMR sails up as a potential 
alternative.
\\ [6mm]
{\bf Acknowledgment} \\ [1mm]
I wish to thank Dr. K{\aa}re Olaussen for making a critical review of
the manuscript.
\\ [6mm]
{\bf References} \\ [2mm]
{\bf [1]} C.M. Will, {\em Theory and Experiment in Gravitational Physics}, 
Cambridge University {\hspace*{7mm}}Press (1993). \\
{\bf [2]} D. {\O}stvang, {\em Doctoral Thesis}, (2001) (gr-qc/0111110). \\
{\bf [3]} M. Castagnino, L. Lara, O. Lombardi, 
{\em Class. Quantum Grav.} {\bf 20}, 369 (2003) \\
{\hspace*{6.2mm}} (quant-ph/0211162). \\
{\bf [4]} D. {\O}stvang, {\em Grav. \& Cosmol.} {\bf 12}, 262 (2006)
(gr-qc/0303107). \\
{\bf [5]} D. {\O}stvang, {\em Grav. \& Cosmol.} {\bf 13}, 1 (2007)
(gr-qc/0201097). \\
{\bf [6]} C.M. Will, K. Nordtvedt, {\em Astrophys. J.} {\bf 177}, 757 (1972). 
\\
{\bf [7]} D. {\O}stvang, {\em Class. Quantum Grav.} {\bf 19}, 4131 (2002)
(gr-qc/9910054). \\
{\bf [8]} J.D. Anderson, P.A. Laing, E.L. Lau, A.S Liu, M.M Nieto, 
S.G. Turyshev, \\
{\hspace*{6.2mm}}{\em Phys. Rev. Lett.} {\bf 81}, 2858 (1998) 
(gr-qc/9808081). \\
{\bf [9]} M. Sethi, A. Batra, D. Lohiya, {\em Phys. Rev.} D {\bf 60}, 108301 
(1999) (astro-ph/9903084). \\
{\bf [10]} D.N. Spergel {\em et al.}, {\em Astrophys. J. Suppl.} {\bf 148}, 
175 (2003) (astro-ph/0302209). \\
{\bf [11]} G. Efstathiou, {\em Mont. Not. R. Astron. Soc.} {\bf 343}, L95 
(2003) (astro-ph/0303127). \\
{\bf [12]} S. Weinberg, {\em Gravitation and Cosmology}, John Wiley ${\&}$ 
Sons, Inc. (1972). \\
{\bf [13]} S. Perlmutter {\em et al.}, {\em ApJ} {\bf 517}, 565 (1999)
(astro-ph/9812133). \\ 
{\bf [14]} A.G. Riess {\em et al.}, {\em AJ} {\bf 116}, 1009 (1998) 
(astro-ph/9805201). \\
{\bf [15]} P. H{\"{o}}flich, K. Nomoto, H. Umeda, J.C. Wheeler,
{\em ApJ} {\bf 528}, 590 (2000) \\
{\hspace*{7.5mm}}(astro-ph/9908226). \\
{\bf [16]} A. Riazuelo, J.-P. Uzan, {\em Phys. Rev.} D {\bf 66}, 
023525 (2002). \\
{\bf [17]} E. Gasta\~{n}aga, E. Garc\'{\i}a-Berro, J. Isern, E. Bravo,
I. Dom\'{\i}nguez, \\
{\hspace*{7.5mm}}{\em Phys. Rev.} D {\bf 65}, 023506 (2002). \\
{\bf [18]} I. Dom\'{\i}nguez, P. H{\"{o}}flich, O. Stranerio, M. Limongi,
A. Chieffi, \\
{\hspace*{7.5mm}} in {\em Proceedings of IAU Colloquium 192} (2003),
Springer Verlag (astro-ph/0311140). 
\end{document}